%% file: main.tex
\documentclass[sigconf]{acmart}

\setcopyright{none}
\renewcommand\footnotetextcopyrightpermission[1]{} 
\usepackage[linesnumbered,algoruled,boxed,lined]{algorithm2e}
\usepackage{algpseudocode}
\usepackage{makecell}
\usepackage{multirow}
\usepackage{booktabs}
\usepackage{hyperref}
\usepackage{cleveref}

\hypersetup{
    colorlinks=true,
    linkcolor=blue,
    filecolor=magenta,
    urlcolor=magenta,
}
\usepackage[all]{hypcap}
\usepackage{booktabs}
\usepackage{hyperref}
\usepackage{url}
\usepackage{xcolor}
\usepackage{tikz}
\usepackage[most]{tcolorbox}

\makeatletter
\g@addto@macro{\@algocf@init}{\SetKwInOut{Parameter}{Parameters}}
\makeatother

\setcopyright{none} 

\begin{document}
\begin{sloppypar}

\title{A Large-scale Fine-grained Analysis of Packages in Open-Source Software Ecosystems}

\author{Xiaoyan Zhou}
\affiliation{
  \institution{Beijing Jiaotong University}
  \country{China}
}
\email{xiaoyanzwendy@gmail.com}

\author{Feiran Liang}
\affiliation{
  \institution{Beijing Jiaotong University}
  \country{China}
}
\email{837504845@qq.com}
   
\author{Zhaojie Xie}
\affiliation{
  \institution{Beijing Jiaotong University}
  \country{China}
}
\email{843193507@qq.com}

\author{Yang Lan}
\affiliation{
  \institution{Beijing Jiaotong University}
  \country{China}
}
\email{18990435388@163.com}

\author{Wenjia Niu}
\affiliation{
  \institution{Beijing Jiaotong University}
  \country{China}
}
\email{niuwj@bjtu.edu.cn}

\author{Jiqiang Liu}
\affiliation{
  \institution{Beijing Jiaotong University}
  \country{China}
}
\email{jqliu@bjtu.edu.cn}
   
\author{Haining Wang}
\affiliation{
  \institution{Virginia Tech}
  \country{America}
}
\email{hnw@vt.ed}

\author{Qiang Li}
\affiliation{
  \institution{Beijing Jiaotong University}
  \country{China}
}
\email{liqiang@bjtu.edu.cn}
\authornote{Corresponding author}

\input{abstract.tex} 
\maketitle

\input{intro.tex}

\input{data.tex}

\input{extract.tex}

\input{R1.tex}

\input{R2.tex}
\input{R3.tex}

\input{R4.tex}

\input{discuss.tex}

\input{relate.tex}

\input{conclusion.tex}
 

\bibliographystyle{ACM-Reference-Format}
\bibliography{ref/ref_software_supply_chain, ref/ref_online, ref/ref_ours.bib, ref/ref_algorithm}
 
\end{sloppypar}
\end{document}

%% file: abstract.tex
\begin{abstract}   
Package managers such as NPM, Maven, and PyPI play a pivotal role in open-source software (OSS) ecosystems, streamlining the distribution and management of various freely available packages.
The fine-grained details within software packages can unveil potential risks within existing OSS ecosystems, offering valuable insights for detecting malicious packages. 
In this study, we undertake a large-scale empirical analysis focusing on fine-grained information (FGI): the metadata, static, and dynamic functions.
Specifically, we investigate the FGI usage across a diverse set of 50,000+ legitimate and 1,000+ malicious packages.
Based on this diverse data collection, we conducted a comparative analysis between legitimate and malicious packages. 
Our findings reveal that 
(1) malicious packages have less metadata content and utilize fewer static and dynamic functions than legitimate ones;
(2) malicious packages demonstrate a higher tendency to invoke HTTP/URL functions as opposed to other application services, such as FTP or SMTP;  
(3) FGI serves as a distinguishable indicator between legitimate and malicious packages; and
(4) one dimension in FGI has sufficient distinguishable capability to detect malicious packages, and combining all dimensions in FGI cannot significantly improve overall performance.

\end{abstract}

%% file: intro.tex
\section{Introduction}

An open-source software (OSS) ecosystem denotes a collection of software projects offering support to developers engaged in application development, encompassing package installation, development, and management.
In software development, developers heavily rely on OSS package managers, such as NPM, PyPI, and RubyGems.
For example, in the construction of a web application, developers turn to Python web frameworks like Django~\cite{django}, Web2py~\cite{web2py}, and Flask~\cite{flask}, leveraging pre-written code to expedite development. 
Notably, 
more than 80\% of the source code of a software product could be from OSS ecosystems~\cite{bavota2013evolution}.
In 2021, the statistics revealed an impressive total of over 2.2 trillion package downloads from OSS ecosystems~\cite{oss_report}.

Regrettably, these reused packages raise security concerns as attackers/hackers can inject malicious codes into packages or compromise benign and legitimate packages to attack systems.
These packages may contain crafted code with malicious intents, such as stealing credentials~\cite{ssc_credential}, installing backdoors~\cite{ssc_backdoor}, and even exploiting computing resources for cryptocurrency mining~\cite{ssc_crypt}.
Recent incidents show that reused packages broke or attacked software across millions of computing platforms. 
For example, in 2018, attackers exploited the development privileges of the `eslint-scope' package to embed malicious executables, compromising numerous systems.

Numerous prior studies~\cite{ma2018constructing, dey2018software, zimmermann2019small, pashchenko2020preliminary} have delved into exploring security concerns within the OSS ecosystem.
\citet{ladisa2023sok} systematically surveyed a large attack surface in the OSS ecosystem. 
Additionally, several approaches~\cite{ohm2020backstabber, alfadel2021empirical, guo2023empirical, SejiaAdria2022Machinelearn, duan2020towards} have been proposed to detect and analyze malicious packages within the OSS ecosystem.
However, the existing malware research has several limitations: a single OSS ecosystem, coarse-grained information, and a lack of comparison between legitimate and malicious packages.
Given the crucial role that software packages play in OSS ecosystems, the security community lacks an understanding of the distinctions between legitimate and malicious packages.
Shedding light on software packages' fine-grained information (FGI) can provide insights into underlying risks in existing OSS ecosystems and enhance malware detection capabilities.

In this paper, we conduct a large-scale empirical study of software packages, investigating 50,000 legitimate and 1,000 malicious packages. 
First, software packages cover 3 OSS ecosystems, including NPM~\cite{ecosrc_npm}, RubyGems~\cite{ecosrc_gem}, and PyPI~\cite{ecosrc_pypi}. 
Second, we provide a comparative analysis between legitimate and malicious packages. 
Third, we explore the fine-grained information (FGI) within software packages: metadata, static, and dynamic functions.
Metadata refers to details about a software package, including its name, version, authors, dependencies, and other pertinent elements. 
Static functions are methods directly integrated into the source code, depending on the programming language employed in the package. 
Dynamic functions are designed to offer flexibility during the installation or runtime phases of the software program. 
These three granularity levels of elements serve as a representation of the package's FGI.

The amalgamation of these diverse and intricate data points enables us to conduct a comprehensive and in-depth comparison between legitimate and malicious packages.
We aim to answer the following research questions (RQs):

\noindent 
\textit{$\bullet$  RQ1: What do legitimate and malicious packages differ at the metadata level?} (Section~\ref{sec:metadata})

\noindent
\textit{$\bullet$  RQ2: What do legitimate and malicious packages differ at the static function level?} (Section~\ref{sec:static})

\noindent
\textit{$\bullet$ RQ3: What do legitimate and malicious packages differ at the dynamic function level? } (Section~\ref{sec:dynamic})

\noindent
\textit{$\bullet$ RQ4: What is the usage of fine-grained information in malware detection?} (Section~\ref{sec:classifier})

\textbf{Findings and Lessons}. 
Our key findings are summarized as follows.
(1) There is a significant difference between legitimate and malicious packages' FGI from the statistical perspective.
Malicious packages have less metadata and employ fewer static/dynamic functions than legitimate packages. 
(2) Static/dynamic functions reflect behavior or operations, so malicious and legitimate packages have different tendencies to call functions.
A noteworthy characteristic of malicious packages is their inclination to use HTTP/URL functions rather than other applications such as FTP or SMTP. 
(3) FGI can be a reliable indicator for distinguishing between legitimate and malicious packages. The detection model based on FGI achieves a promising performance, with an accuracy of 97.5\% and a recall of 94.4\%.
(4) Simply combining all FGI elements only slightly improves the overall malware detection performance because each FGI dimension has distinguishable capability.

\noindent
\textbf{Roadmap}. 
The remainder of this paper is organized as follows.
Section 2 presents the fine-grained information for legitimate and malicious packages.
Section 3 details the difference at the metadata level.
Section 4 presents the difference at the static function level.
Section 5 illustrates the difference at the dynamic function level.
Section 6 presents a malware detection method based on FGI.
Section 7 discusses the limitations. 
Section 8 surveys related work, and finally, Section 9 concludes.

%% file: data.tex
\section{Fine-grained Information}

\begin{table}[bht] \small
	\caption{The number of software packages.}
	\centering
	\begin{tabular}{c c c c }
		\toprule
		OSS Ecosystem                   &  Language        &  Legitimate pkg.           &  Malicious pkg.\\
		\toprule
		NPM~\cite{ecosrc_npm}          &JavaScript    &17,728      &  686       \\
		
		PyPi~\cite{ecosrc_pypi}        &  Python   &    17,011          &       259 \\
		
		RubyGems~\cite{ecosrc_gem}    & Ruby      & 15,397           &  43         \\
		
		\toprule
	\end{tabular}
    \label{tab:data}
\end{table}

\subsection{Data Collection Methodology }

To explore the software's FGI, we need to gather legitimate/malicious packages. 
Given this goal, our investigation focused on accessible packages.

\textbf{Legitimate Packages}.
Our data collection centered on three OSS ecosystems, including NPM~\cite{ecosrc_npm}, PyPI~\cite{ecosrc_pypi}, and RubyGems~\cite{ecosrc_gem}. 
They are popular package-management systems that automate the installation, upgrading, configuration, and removal of software packages.
We relied on three OSS ecosystems to collect publicly disclosed legitimate packages, where each package is associated with a unique name and version.
In particular, we utilized a web crawler to download software packages from the OSS ecosystems.
While there exist other OSS ecosystems, such as Composer~\cite{ecosrc_packagist}, NuGet~\cite{ecosrc_nuget}, and Maven~\cite{ecosrc_maven}, we focused on the three mentioned ecosystems as they are: 
(1) public, free, and easily accessible via API querying, allowing for reproducibility and follow-on studies, 
(2) the most widely used programming languages, e.g., JavaScript and Python, 
and (3) manually vetted and curated by package administrators.

\textbf{Malicious Packages}.
Our data collection is centered on public and disclosed malicious packages. A malicious package represents a combination of harmful software components designed to threaten the functionality and security of a computer system.
We leverage three sources: the GitHub Security Advisory Database~\cite{github-advisory}, Backstabber-Knife~\cite{backstabbers-online} dataset, and MalOSS~\cite{duan2020towards}. 
The GitHub Security Advisory Database~\cite{github-advisory} is a free and open-source repository of security advisories, where we downloaded the corresponding malicious packages with versions from it.
Backstabber-Knife~\cite{backstabbers-online} is a collection of malicious packages against OSS ecosystems, yet it is not publicly available for download.
We search those package names in registry mirrors~\cite{ruby_aliyun} \cite{npm_aliyun} \cite{pypi_tsinghua} \cite{ruby_tsinghua} \cite{pypi_ustc}  \cite{npm_ustc}.
If a malicious package exists, we download it from the registry mirrors; otherwise, we skip it.
MalOSS~\cite{duan2020towards} is a private open-source repository, and we downloaded all available malicious packages from it.  
One typical problem is that many malicious packages from different sources may be duplicated. 
We use a heuristic rule to remove the duplicated one: if two packages have the same name and version, they belong to the same packages. 

We have collected 50,000 legitimate and 1,000 malicious packages. 
Table~\ref{tab:data} lists the data distribution of software packages, with NPM comprising 17,728 legitimate packages and 686 malicious packages, PyPI containing 17,011 legitimate packages and 259 malicious packages, and RubyGems including 15,397 legitimate packages and 43 malicious packages.

%% file: extract.tex
\subsection{Fine-Grained Information (FGI) of Packages }

\begin{table}[!t]\small
    \caption{The difference between a malicious package and a legitimate package.}
    \label{tab:difference:example}
    \centering
    \begin{tabular}{l c c}
    \toprule
                 &  Legitimate  &  Malware   \\
    \toprule
    Name            & loglib & loglib-modules           \\
    Description  &  \makecell[l]{A decent logging system\\
                    with some settings built in...... } & None          \\
    
    Author  & \makecell[l]{Logan Houston \\ 
                    houston4509@gmail.com} & ALou3   \\
    
    Homepage&   Github URL & None    \\
    
    Dependency    &    neotermcolor (>=2.0) & None          \\ \hline
    
    \makecell[c]{Static \\ function} & \makecell[c]{process execution \\ 
                                            fetches data over the network \\
                                            read/write files and dir     } &             
                                             \makecell[c]{ reads hidden code\\ reads files and dir}\\ \hline
    
     \makecell[c]{Dynamic \\ function} & \makecell[c]{data sent, file write\&read, \\ 
                                        new dir, permissions, \\
                                        file path, process execution} &  \makecell[c]{file read\\ process execution}\\
     \toprule
    \end{tabular}

\end{table}

Attackers and criminals purposefully craft malicious packages to conduct malicious behaviors, e.g., stealing private information and disrupting systems. 
Legitimate packages in OSS ecosystems provide software users with certain functionalities, such as resource management, communications, data access, and user interface creation.
Hence, there are significant differences between malicious and legitimate packages.

\textit{Take one example}. 
Table~\ref{tab:difference:example} lists the differences between a malicious package `loglib-modules' and a legitimate package `loglib'.
The `loglib' is a legitimate software package, and the `loglib-modules' is a malicious package that deceives users into downloading it. 
At the package metadata level, `loglib' has comprehensive and detailed information, while `loglib-modules' lacks most of it. 
For instance, `loglib-modules' has an empty description, a Null homepage, and no dependency.
At the static function level, `loglib' has more operations than `loglib-modules'. 
At the dynamic function level,  `loglib' has more operations than `loglib-modules' in the logging file.

We provide the FGI of the OSS package for analyzing software packages and defense techniques, as shown in Figure~\ref{fig:fgi}.
Package metadata is the coarse-grained information stored in the configuration file, essential for package management, software distribution, and system administration.
The information at this level provides the package fundamentals, such as its name, version, dependencies, license, and other relevant attributes. 
A static function is the fine-grained information stored in the source code file. To obtain a static function, we need to download and unpack the package and parse its source code files.
The dynamic function is the fine-grained information of a software package.
We must run the software package and record dynamic functions during its execution.
The information at the dynamic level is challenging to obtain and analyze because software execution relies on dependency libraries and operating systems.
In this work, we extract the FGI of software packages and comprehensively analyze software packages to answer search questions.

\subsection{Threats to FGI's Validity}

\textbf{Threats to Dataset Size}. 
One concern is that the dataset size may not be large enough to represent the comprehensive pattern of legitimate and malicious packages. 
Our dataset only contains 50,000+ legitimate packages and 1,000 malicious packages, while the entire OSS ecosystem may have millions of packages. There are two reasons for limiting the dataset size.
First, static and dynamic functions need to be extracted from the software package's source code and runtime behavior, leading to a high time cost and manual effort.
Second, the number of available malicious packages is limited. 
So far, the data collection of the OSS malicious package is still in its infancy, and accessible/downloadable malicious packages are limited due to ethical and legal considerations.

\textbf{Threats to FGI Extraction}.
Another concern is that the FGI extraction has a performance issue, where its accuracy is low and unacceptable in practice.
In our study, the metadata extraction achieves 100\% accuracy, and the static function extraction achieves 99\% accuracy.
The dynamic function extraction has a relatively lower success rate, limited to Unix-like operating systems. 
We acknowledge that our study's FGI extraction relies on existing tools or approaches rather than our proposed tools or algorithms.
In the future, we will integrate more cutting-edge tools to improve the performance of the FGI extraction.

%% file: R1.tex

\section{RQ1: FGI at the Metadata}
\label{sec:metadata}

This section provides the metadata analysis for comparing the legitimate and malicious packages, with 50,000 legitimate and 1,000 malicious packages, listed in Table~\ref{tab:data}.
We divide the 50,000 legitimate packages into popular and random software packages. 
The popular category includes the most downloaded or highest Pagerank packages, while the random category contains packages randomly selected from three OSS ecosystems.

\subsection{Extracting Package Metadata}

\begin{figure}[!t]
    \centering
    \includegraphics[width=2.9 in]{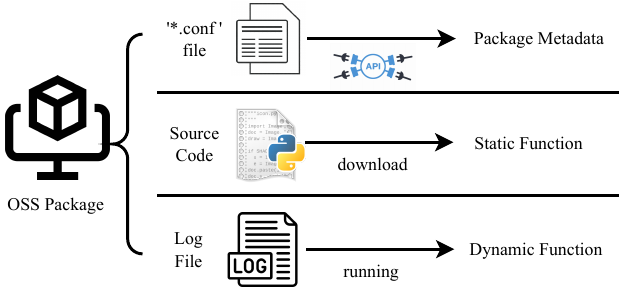}
    \caption{The OSS package extraction at the FGI level.}
    \label{fig:fgi}
\end{figure}

Given a software package, we extract its metadata and store it as the key-value form. 
Metadata constitutes descriptive information or data that provides additional context, characteristics, or attributes about an OSS package. 
There are typically two methods for obtaining package metadata: querying APIs and parsing configuration files.
For APIs, we use the package manager's API to fetch metadata for packages automatically.
For configuration files, we need to download the package and then extract the metadata from the configuration file.
Specifically, common formats of metadata files include JSON, XML, YAML, and INI files.

We provide essential information in the metadata as follows. 
(1) Package name is an identifier in the SSC ecosystem. 
(2) Package version helps with package management and versioning control. 
(3) Description presents the purpose or function of the package.
(4) The author's name is the individual or group responsible for maintenance.
(5) The homepage provides the URL of the package's official website or homepage, offering additional information and documentation.
(6) Dependency $\&$ Dependents refer to other packages that this package relies on. 
For example, if the package $p_i$ reuses the function from the package $p_j$.
In this case, the $p_j$ is the dependency package of the $p_i$, and the $p_i$ is the  $p_j$'s dependent package.

\subsection{Findings and Lessons}

\begin{figure}[!t]
	\centering
	\includegraphics[width=2.4in]{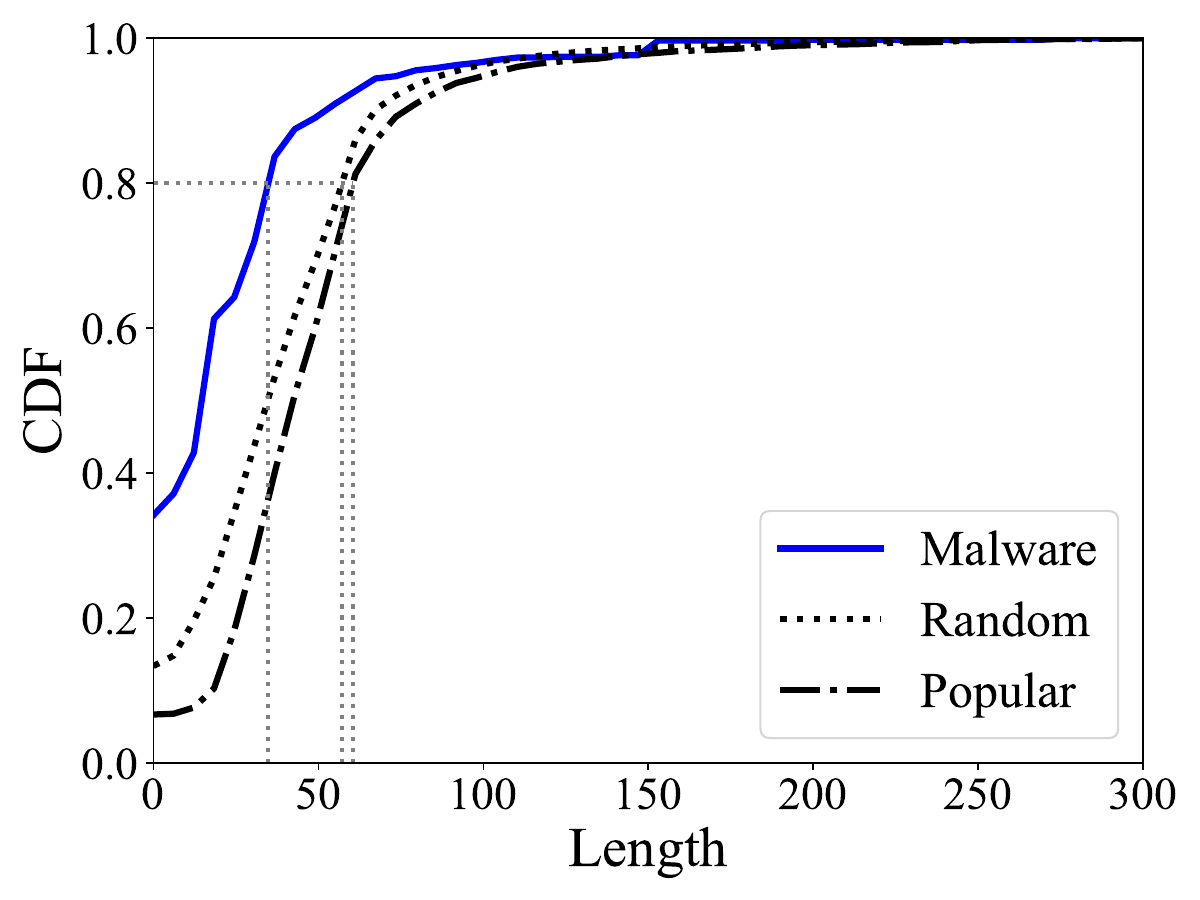}
	\caption{The CDF of the package description length.}
    \label{fig:pack:des}
\end{figure}

\textit{Package Description}.
When software developers release a package, they typically embed or hardcoded a description, such as a README document. 
This description offers valuable insights into the package's functionality and purpose. 
We quantify the richness of this description by measuring its string length. 
In Figure~\ref{fig:pack:des}, we present a cumulative distribution function (CDF) plot that outlines the distribution of package description lengths.
The outcomes reveal that 80\% of malicious packages have descriptions comprising fewer than 40 words, and approximately 37\% of these malicious packages lack any descriptive content altogether.
We discovered two common patterns when examining malicious packages with descriptions over 200 characters. 
Some malicious packages replicate the descriptions of legitimate counterparts (e.g., `@employee-experience/common'), while others use descriptions to document tracking functionality, as seen in `colourama-0.1.6' and `yiffparty-0.04'.
By contrast, legitimate packages (both popular and random) have longer descriptions than malicious packages.
Additionally, we observed that legitimate software packages provide the  package's description metadata field and extensively describe the package's purpose and additional details.

\begin{figure}[!t]
	\centering
	\includegraphics[width=2.4in]{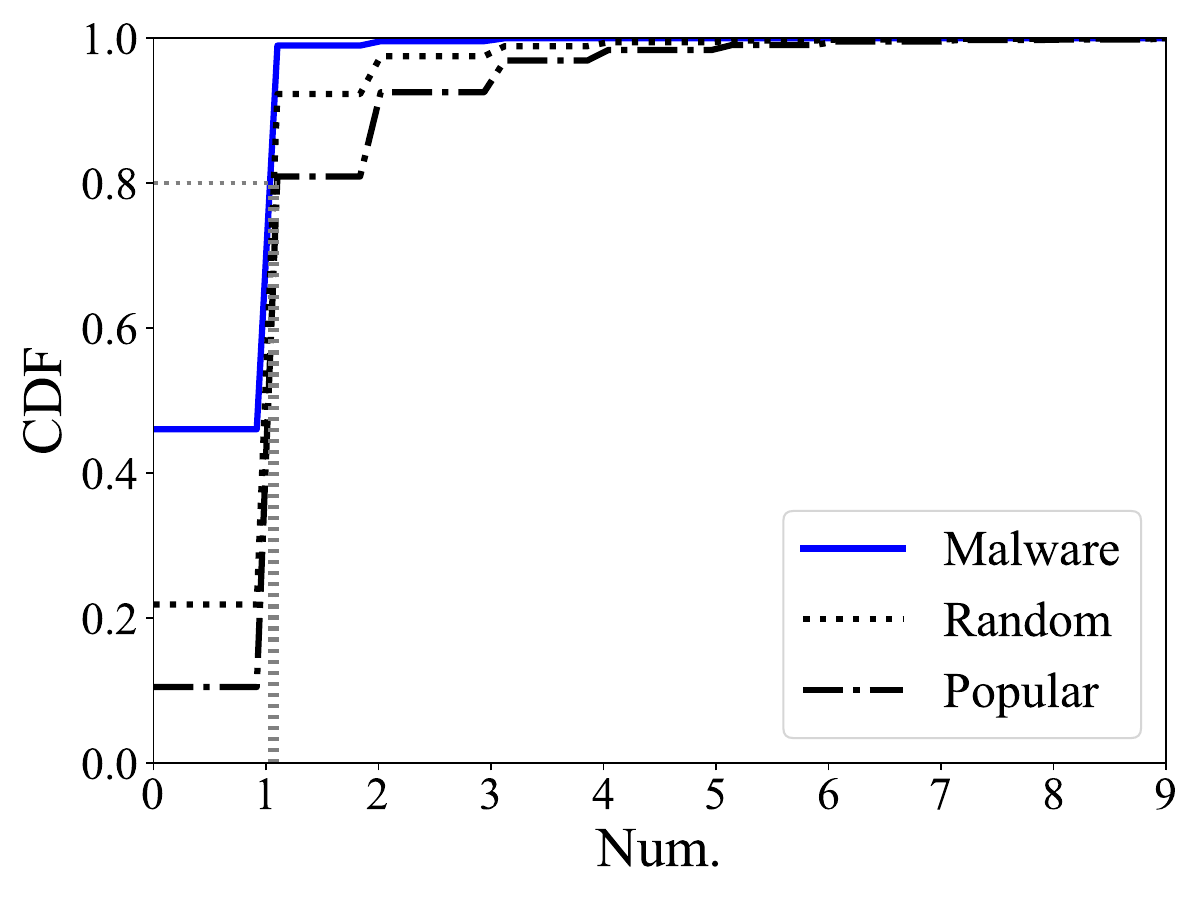}
	\caption{The number of authors/maintainers for software packages.}
    \label{fig:author}
\end{figure}

\textit{Author and Maintainer}.
We conduct an in-depth analysis of the number of authors and maintainers associated with software packages.
Figure~\ref{fig:author} depicts the CDF of the authors/maintainer number per package.
A substantial proportion of malicious packages demonstrate few authors and maintainers. 
Specifically, nearly half of the malicious packages lack any listed authors or maintainers, and approximately 80\% of these packages are associated with only a single author or maintainer.
The reason is that attackers prefer maintaining a discreet online presence within OSS ecosystems. 
The inherent implication is that multiple malicious packages may point to the same attacker. 
Only 4 malicious packages have an author and a maintainer, e.g., `python-dateutils' and `get-text'.
We manually inspect the activity status of authors or maintainers from malicious packages.  
The author accounts linked to malicious packages are no longer active. 
In contrast, 80\% of popular packages have 4 authors/maintainers.
Legitimate random packages also have a consistent distribution, with 80\% of them having two or more authors/maintainers.
In addition, we observed some popular software packages (5.2\%, 1,160/22,314) have hundreds of authors/maintainers., e.g., `lodash' and `chalk'.
These packages belong to a large open-source project with hundreds of distributions, leading to many maintainers.

\begin{figure}[!t]
	\centering
	\includegraphics[width=3in]{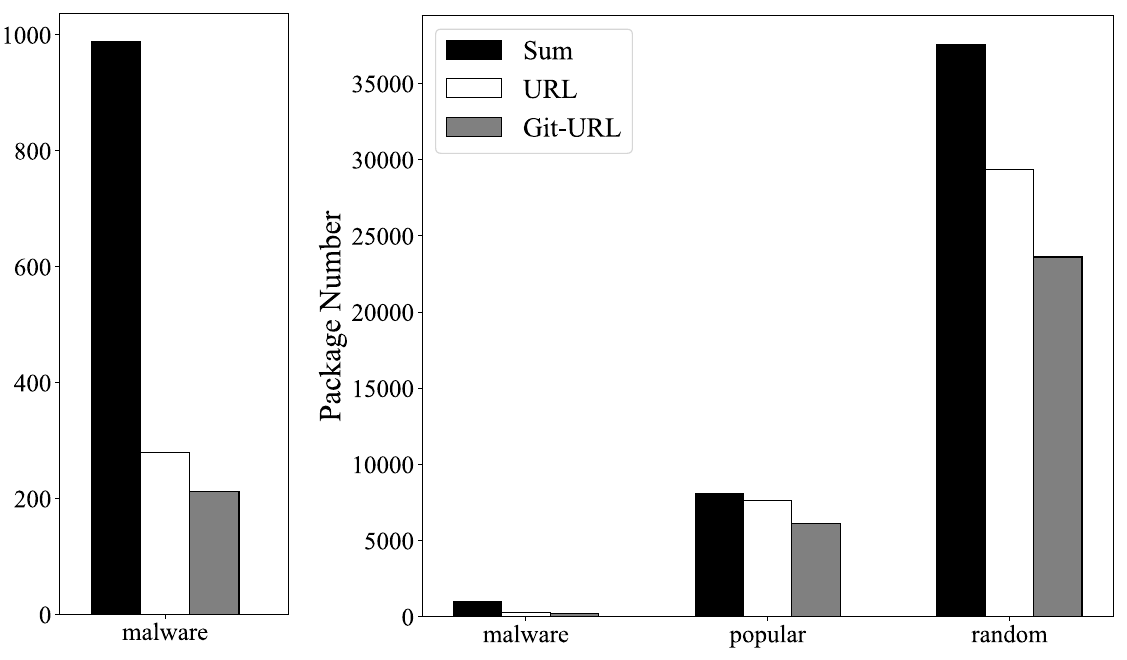}
	\caption{The distribution of URL from software packages.}
    \label{fig:url}
\end{figure}

\textit{HomePage and Code Repository}.
The homepage of a software package serves as the official website, containing details about the program's utility, installation instructions, tutorials, and related wiki sites. 
The code repository is used to store the source code of the software package. 
In this case, we use GitHub as the open-source software code repository. 
GitHub is the world's largest online source code hosting platform, with nearly 60 million users and 190 million open-source code libraries.
Our analysis uses the keyword `git' to match URLs extracted from packages. 
If a URL contains the string 'git,' it indicates that the package utilizes GitHub as its code repository.
Figure~\ref{fig:url} depicts the distribution of homepages and code repositories across the dataset of 50,000 packages.
For legitimate packages, a substantial majority, comprising 45,668 software packages (91.4\%), include the homepage and the code repository URLs. 
A remarkable 80\% of all URLs are linked to GitHub, highlighting its dominant position as the preferred choice within the OSS ecosystem. 
The remaining 20\% of URLs primarily connect to social media platforms such as Twitter and Facebook.
Conversely, malicious packages exhibit a starkly different pattern. 
Only 292 software packages (29.6\%) contain a URL link, and 213 packages (21.6\%) have a GitHub link.
Adversaries do not explicitly provide the homepage or code repository of the malicious packages. 
We further inspect those URLs from malicious packages. 
Several malicious packages simulate their URLs using public websites like `https://www.google.com'. Moreover, deceptive URLs are prevalent, as evidenced by instances like `http://pypipack@protonmail.com' and `https://example.com'. 
Surprisingly, we find the code repositories for these malicious packages within the Git URL, indeed hosting the source codes and relative package versions, which seems counter-intuitive.

\begin{figure}[!t]
	\centering
	\includegraphics[width=2.4in]{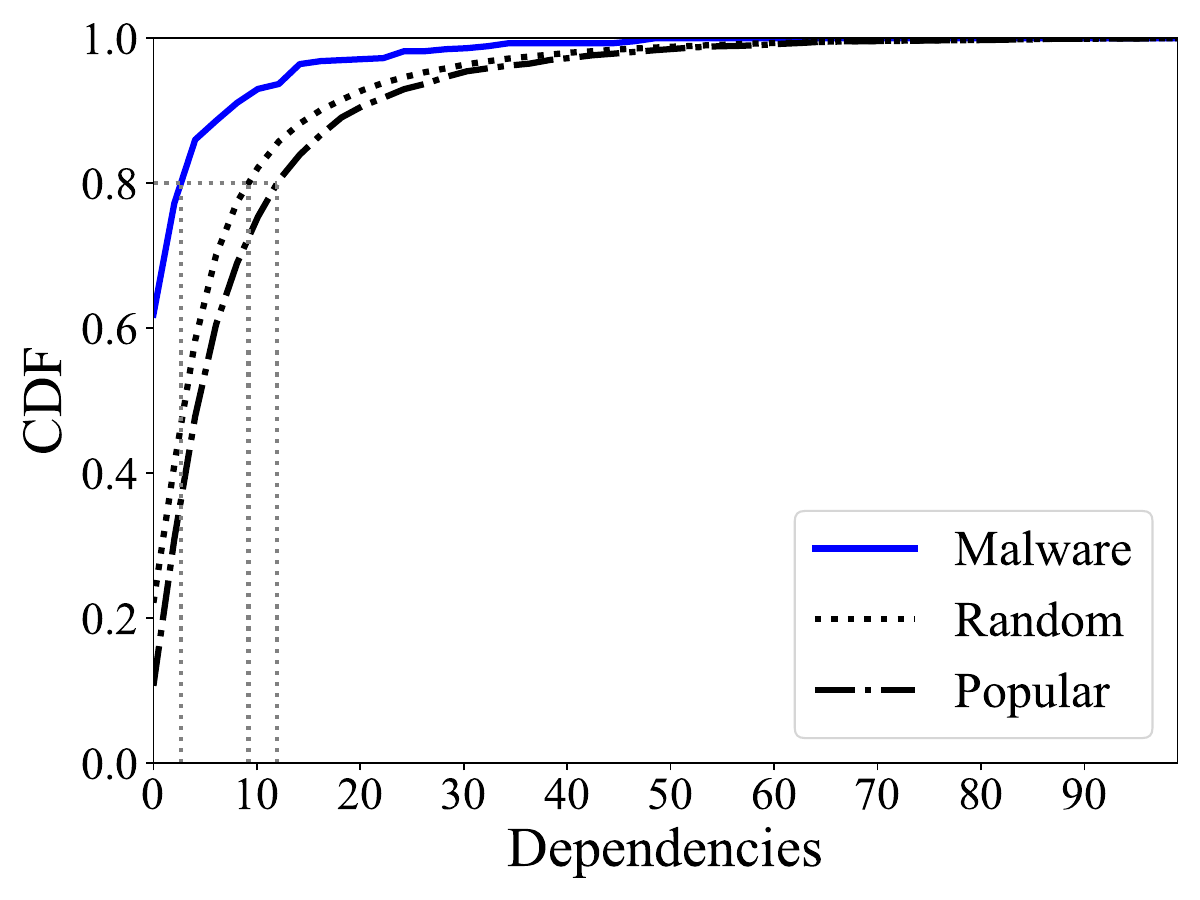}
	\caption{The CDF of dependencies of software packages.}
    \label{fig:dep}
\end{figure}

\textit{Dependencies}.
Our study comprehensively compares dependencies among legitimate and malicious packages within three distinct OSS ecosystems.
Figure~\ref{fig:dep} depicts a CDF plot of the number of dependencies per package. 
This statistical analysis reveals a distinct pattern: approximately 80\% of malicious packages have fewer than 3 dependencies, whereas 80\% of legitimate packages exhibit more than 10 dependencies. 
Further, nearly 60\% of malicious packages lack any dependencies. 
This indicates that the number of dependencies in malicious packages is lower than in legitimate software packages. 
Malicious packages have fewer features and functions than legitimate ones, leading to less reuse of third-party libraries.

\begin{figure}[!t]
    \centering
    \includegraphics[width=3.1 in]{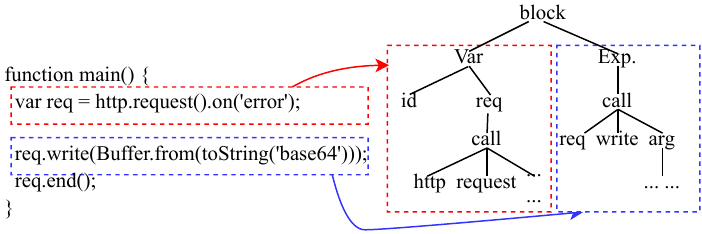}
    \caption{ Source code file and the corresponding AST.}
    \label{fig:back:ast}
\end{figure}

\begin{figure*}[!t]
    \begin{tabular}{ c   c  c }
    \begin{minipage}[t]{0.31\linewidth}
    \centering
    \includegraphics[width=1.0\linewidth]{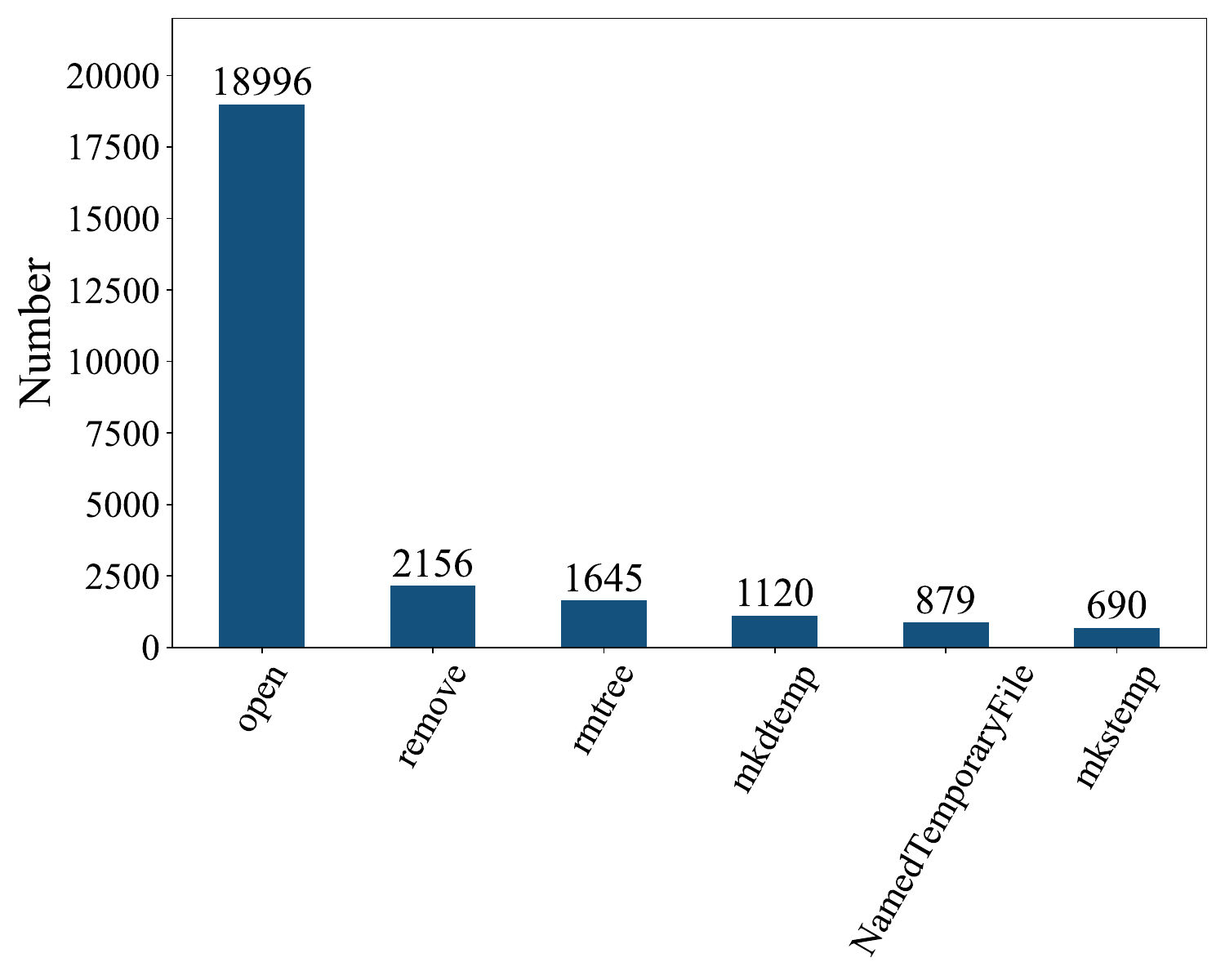}
	\caption{$S_{same}$: File-related functions.}
	\label{fig:file:api}
    \end{minipage}
    &
    \begin{minipage}[t]{0.31\linewidth}
    \includegraphics[width=1.0\linewidth]{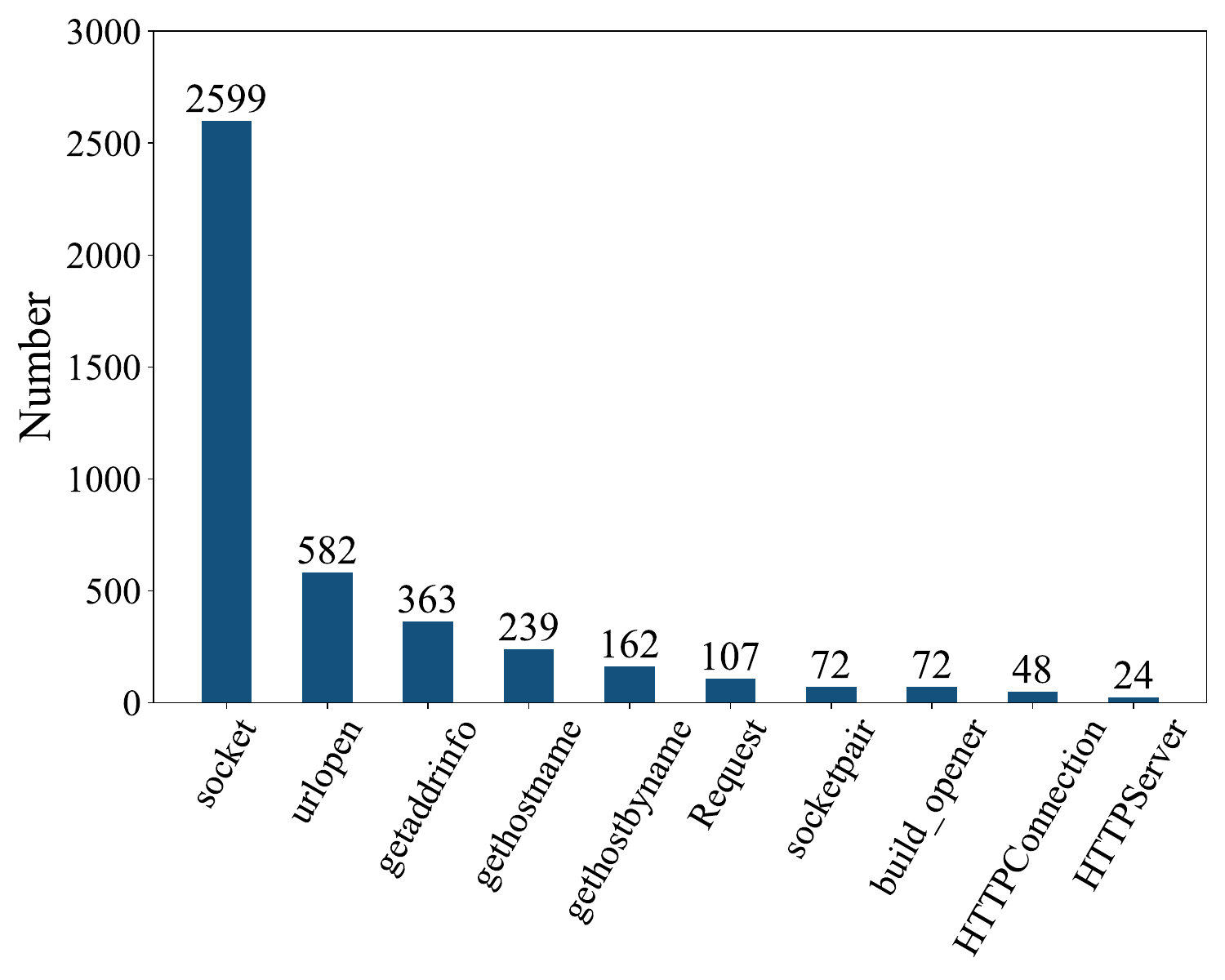}
	\caption{$S_{same}$: Network-related functions.}
	\label{fig:net:api}
    \end{minipage}
	&
    \begin{minipage}[t]{0.31\linewidth}
    \includegraphics[width=1.0\linewidth]{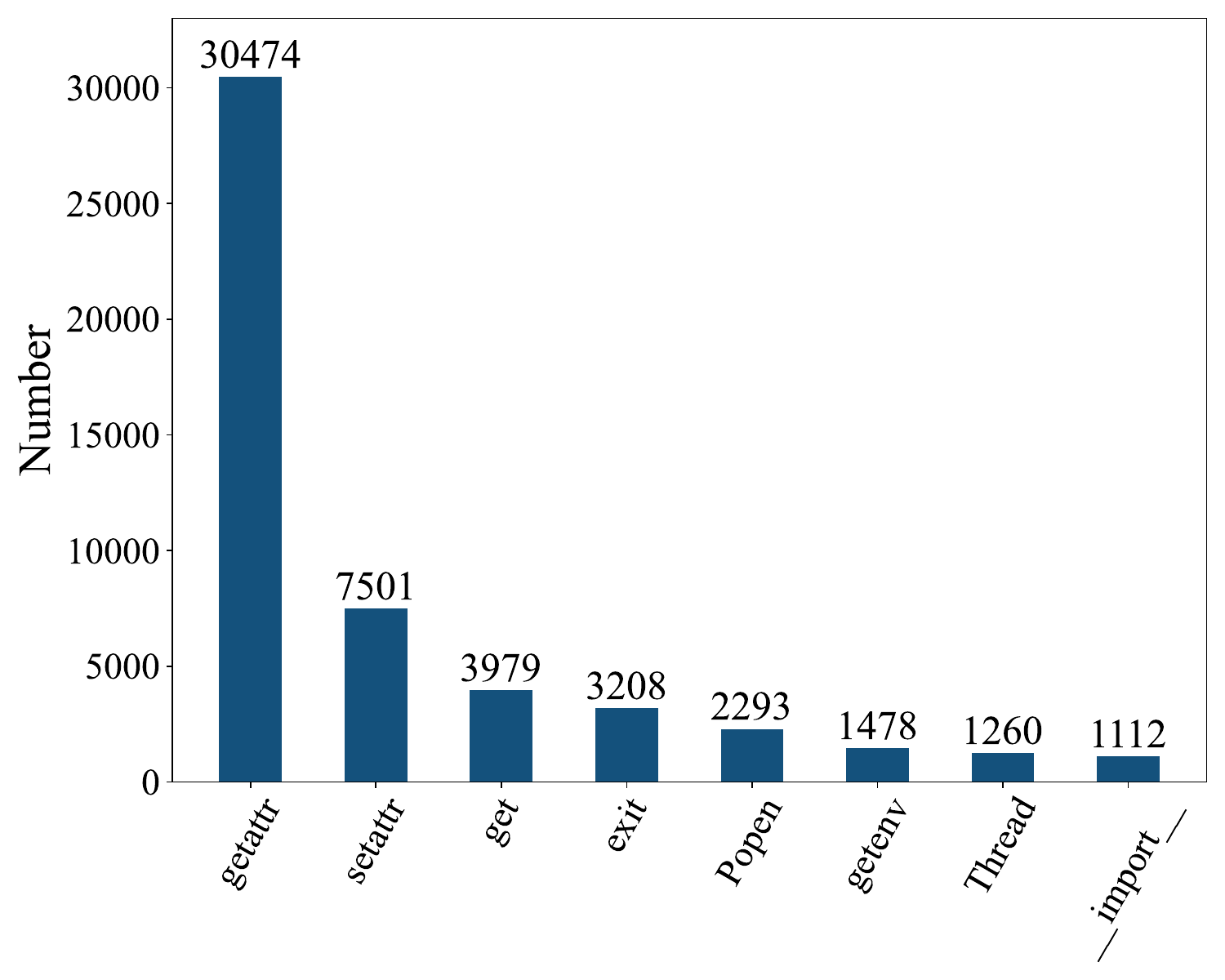}
	\caption{$S_{same}$: process-related functions.}
	\label{fig:pro:api}
    \end{minipage}\\
    \begin{minipage}[t]{0.31\linewidth}
        \centering
       \includegraphics[width=1.0\linewidth]{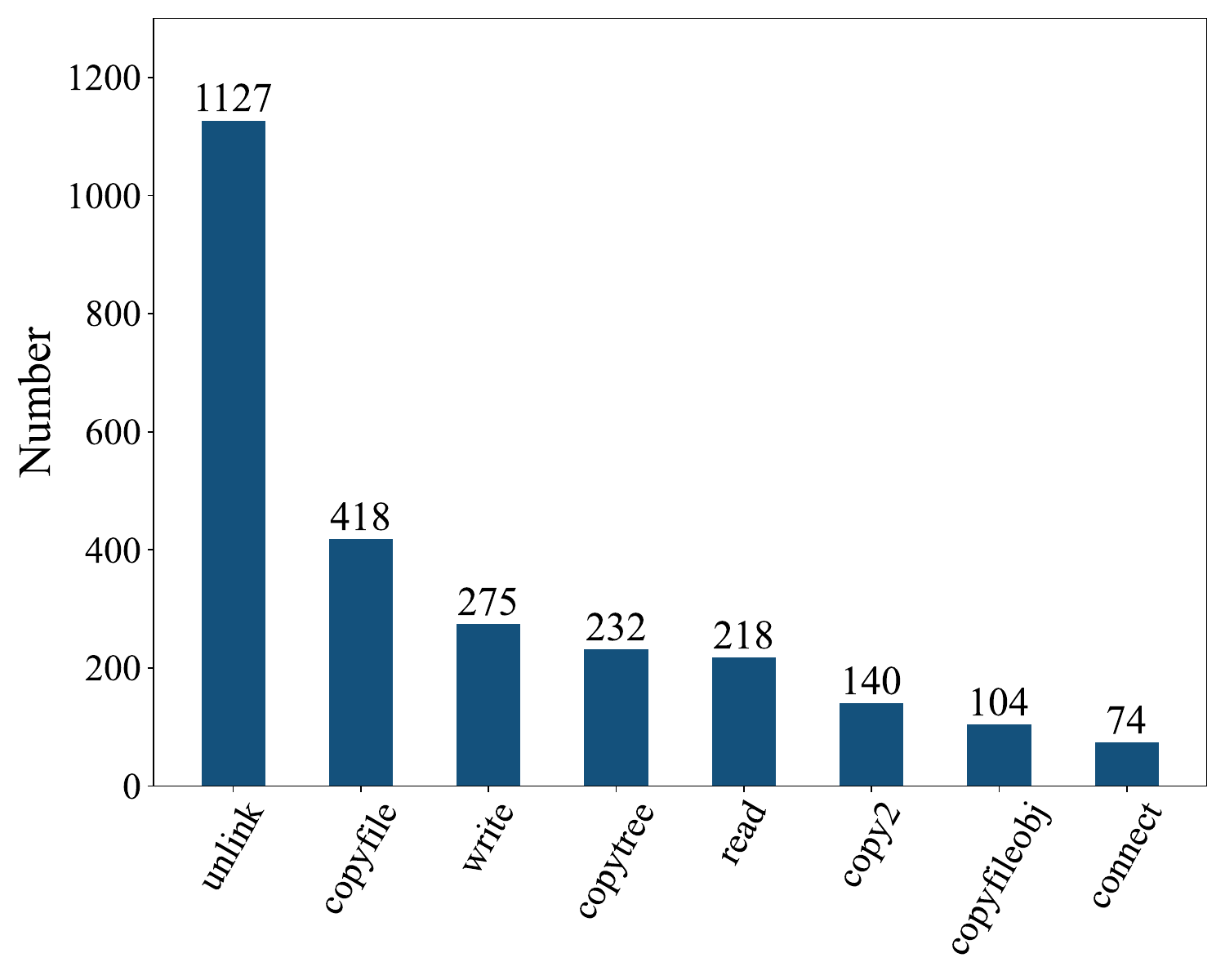}
        \caption{$S_{P_r^-}$: File-related functions.}
        \label{fig:file:api1}
        \end{minipage}
        &  \begin{minipage}[t]{0.31\linewidth}
        \includegraphics[width=1.0\linewidth]{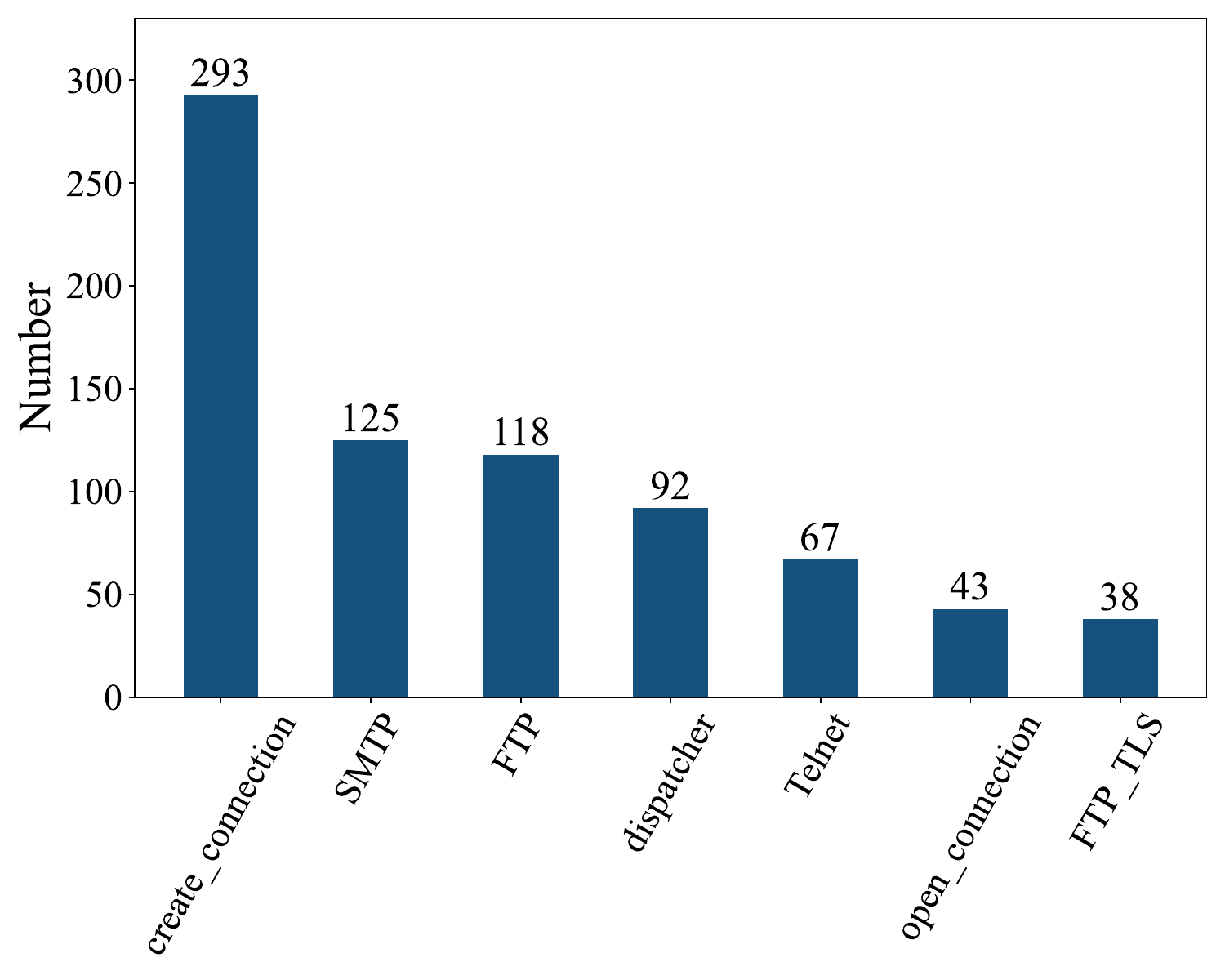}
        \caption{$S_{P_r^-}$: Network-related functions.}
        \label{fig:net:api1}
        \end{minipage}       &
        \begin{minipage}[t]{0.31\linewidth}
        \includegraphics[width=1.0\linewidth]{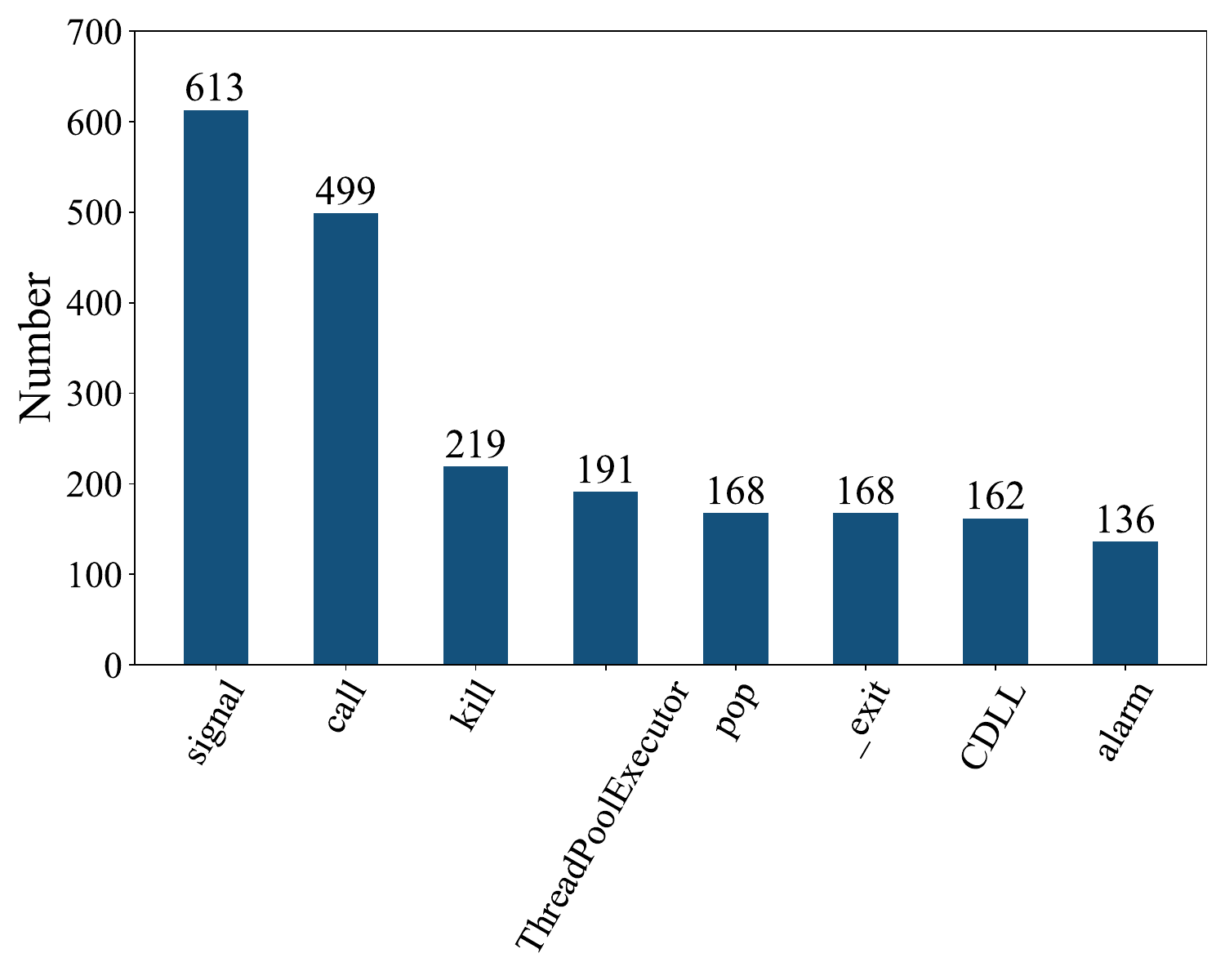}
        \caption{$S_{P_r^-}$: Process-related functions.}
        \label{fig:pro:api1}
        \end{minipage}
    \end{tabular}
\end{figure*}

\textbf{Lessons} learned are as follows.
(1) The fine-grained information at the metadata level has a distinguished pattern or feature to detect malicious packages from legitimate ones, e.g., shorter package descriptions, fewer author numbers, missing URLs, and fewer dependencies.
This characteristic holds promise in distinguishing between malicious and legitimate packages.
(2) The drawback is that the metadata would be immediately invalid once the attackers become aware of them, as they are easy to craft (e.g., number of authors). Thus,
detection tools are not recommended to rely heavily on metadata to detect malicious packages.

%% file: R2.tex

\section{RQ2: FGI in Static Functions}
\label{sec:static}

This section provides the static function analysis for comparing the legitimate and malicious packages.
Note that static functions need to download packages, unpack them into files, and build ASTs.  
Hence, we only pick 1,905 legitimate and 259 malicious packages in the PyPI ecosystem, as listed in Table~\ref{tab:data}.

\subsection{Extracting Static Function}

To obtain static functions, we need to download the software package and unpack it to a folder.
Unpacking refers to extracting a package's contents from a compressed or archived format into usable files and folders. 
This operation involves various tasks, such as decompression, file extraction, and directory creation. 
After unpacking a software package, we obtain all its files, including source code files, resource files, and binary files. 
Static functions stay in the source code files.
For example, the source code file in the PyPI ecosystem uses the extension `.py', while that in the NPM ecosystem uses the extension `.js'.

We convert each source-code file into an abstract syntax tree (AST). 
An AST is a tree representation of the abstract syntax block call structure of code in the compilation and decompilation process. 
A parser automatically generates it from a source code file based on the code's syntactic structure.
Different AST subtrees represent various code snippets in the source code. 
Figure\ref{fig:back:ast} shows an AST corresponding to the source code file, where the left side displays the source code, and the right side depicts the AST. 
The code snapshot is extracted from the `build.js' file in the `teams-data' package. 
The lines connecting the source code and AST in Figure~\ref{fig:back:ast} indicate that a node in the AST corresponds to an expression or a function in the source code. 
The nodes in an AST contain the following forms: expressions, statements, declarations, function names, parameters, and types. 
We traverse the AST through the depth-first search.

We use the regex matching to identify static functions in the AST. 
Table~\ref{tab:function} shows the overall information of static functions from 3 programming language documents.
Specifically, we extract the three categories of static functions, including network-related, file-related, and process-related functions. 
If an AST node matches a function name, we extract the relevant information, including functions, parameters, and matched files.

\begin{table}[!t] \small
    \caption{The static/dynamic function list. }
    \label{tab:function}
    \centering
    \begin{tabular}{  c c  }
    \toprule
    Category                              &  Number (Programming Language) \\
    \toprule
    \multirow{1}*{Network-related Function }      &    168 (Python) + 12(JavaScript)+ 88(Ruby)   \\
    \multirow{1}*{File-related Function }          &    114 (Python) + 39(JavaScript)+ 89(Ruby)   \\
    \multirow{1}*{Process-related Function }       &    72 (Python) + 15(JavaScript)+ 1(Ruby)   \\

   \toprule    
    \end{tabular}
    
\end{table}

\subsection{Findings and Lessons}

We use the $P_m$ to represent the malicious package set and $P_r$ to present the legitimate package set, as follows.
$$P_m = \{p_m^1, \dots, p_m^i, \dots, p_m^{n_1}\}$$
$$P_r = \{p_m^1, \dots, p_r^j, \dots, p_m^{n_2}\}$$
where $p_m^i$ is the $i$th malicious package, $p_r^j$ is the $j$th legitimate package, $n_1$ represents the number of malicious packages, and $n_2$ represents the number of legitimate software.
For each software package, we extract its corresponding set of static functions, denoted as $p^i$=\{$f_1^i$, $\dots$, $f_k^{i}$\}, where $f_1^i$ represents the first function called by the $i$ th package.
Hence, the set ($P_m$/$P_r$) can be converted into the function set, where $n$ represents the total number of malicious package static functions and $n'$ represents the total number of legitimate package static functions. 
$$P_m = \{f_m^1, \dots, f_m^j, \dots, f_m^{n}\}$$
$$P_r = \{f_r^1, \dots, f_r^j, \dots, f_r^{n'}\}$$
If a function appears in legitimate and malicious packages, we store it in the set 
$S_{same}$ = $\{P_m\wedge P_r\}$.
We use the set $S_{P_r^-}$ = $\{P_r - P_m\}$ to represent functions that only appear in the legitimate package, and the set $S_{P_m^-}$ = $\{P_m - P_r\}$ to represent functions that only appear in the malicious package.
Each function falls into three categories (Table~\ref{tab:function}): network-related, file-related, and process-related functions. 
We explore all packages and generate those function sets.

\textit{File-related Functions}.
Figure~\ref{fig:file:api} shows the distribution of file-related functions in the set $S_{same}$, and Figure~\ref{fig:file:api1} depicts the distribution of file-related functions in the set $S_{P_r^-}$.
The X-axis is the function name, and the Y-axis is the number of the software package.
In the set $S_{same}$, we find 15 functions, where the most frequent one is `open', followed by `remove', `rmtree', and `mkdtemp'.
On average, the function `open' has nearly 8 function calls in the set $S_{same}$ per package.
In the set $S_{P_r^-}$, there are 26 file-related functions, where the most frequent one is `unlink', followed by `copyfile' and `write'.
It seems anti-intuition that malicious packages only call ``open'' functions rather than ``write'' and ``read'' functions. 
The plausible reason is that the malicious package lacks permissions to limit the ``write'' and ``read'' functions.

\textit{Network-related Functions}.
Figure~\ref{fig:net:api} describes the distribution of network-related functions in the set $S_{same}$, and Figure~\ref{fig:net:api1} depicts the distribution of network-related functions in the set $S_{P_r^-}$.
In the set $S_{same}$, there are a total of 10 network-related functions. 
The function  `socket' is the most frequent, nearly 2,599 times, followed by `URLopen' and `getaddrinfo'. 
We find that both malicious and legitimate packages are prone to utilize the functions of HTTP or URL operations, e.g., `HTTPConnection', `Request', and `URLopen'. 
By contrast, the `SMTP' and `FTP' appear in the set $S_{P_r^-}$, representing that malicious packages rarely use those functions. 
Combining Figure~\ref{fig:net:api} and Figure~\ref{fig:net:api1}, we find that malicious packages are prone to leverage HTTP-related operations instead of other application services. 
We believe this observation is consistent with many malicious packages originating from command-and-control (C\&C) servers. 
The C\&C servers, controlled by attackers or cybercriminals, receive commands from and send commands to the malware-compromised system of the target. 
This communication involves sockets, HTTP requests, and HTTP connections.

\textit{Process-related Functions}.
We further provide an analysis of the process-related functions in the malicious and legitimate packages. 
Figure~\ref{fig:pro:api} displays the distribution of process-related functions in the set $S_{same}$. 
We find that 72 functions appear in malicious and legitimate packages. The most frequent one is `getattr', followed by `setattr' and `get'.
Figure~\ref{fig:pro:api1} shows the distribution of process-related functions in the set $S_{P_r^-}$. 
The process operation has 72 functions, the most frequent of which is `signal', followed by `call' and `kill'. 
Combining Figure~\ref{fig:pro:api} and Figure~\ref{fig:pro:api1}, we find that the malicious packages do not use functions related to OS scheduling and services.
By contrast, process-related functions in malware usually correlate with file-related or network-related operations.

\textit{Unique Functions in the Malware}.
The set $S_{P_m^-}$ represents functions only the malicious packages use, but the legitimate ones do not. 
We find that the set $S_{-P_m}$ is equal to $\emptyset$. 
The result illustrates that there is no static function in legitimate packages, only in malicious ones. 
The plausible reason is that adversaries develop a malicious package by using general functions to fulfill their malicious intent.

\textbf{Lessons} learned are as follows.
(1) The set $S_{same}$ cannot be a pivotal discriminator for malicious packages, and $S_{r^-}$/$S_{m^-}$ can act as a distinguished indicator. 
Yet, the $S_{m^-}$ is an empty set.
(2) Malicious packages demonstrate a higher tendency to invoke HTTP/socket functions as opposed to other application services, such as FTP, SMTP, and Telnet; they also correlate with file-related or network-related operations.
(3) Static functions reflect the behavior of malicious code, e.g., stealing sensitive data and exfiltrating it to the attacker's server via an HTTP GET request.

%% file: R3.tex

\section{RQ3: FGI in Dynamic Functions}
\label{sec:dynamic}

This section provides a dynamic function analysis that compares legitimate and malicious packages.
Note that dynamic functions need to download packages, unarchive them into files, run them in the sandbox, and record logging files.  
Hence, we picked only 1,822 legitimate and 686 malicious packages in the NPM ecosystem and 1,900 legitimate and 43 malicious packages in the RubyGems ecosystem.

\begin{figure}[!t]
    \centering
    \includegraphics[width=2.9 in]{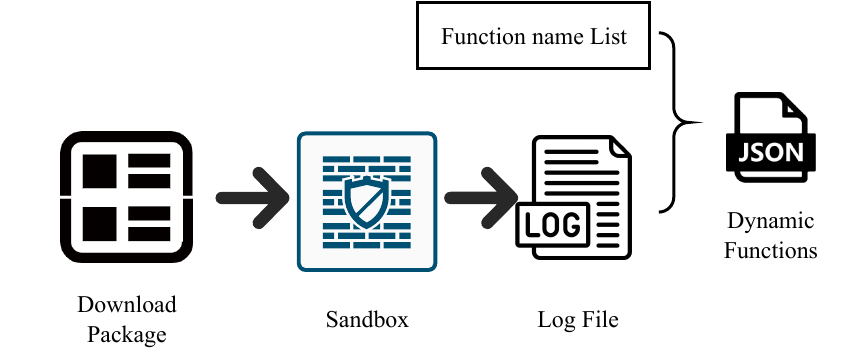}
    \caption{Extracting dynamic functions of packages.}
    \label{fig:ext:dynamic}
\end{figure}

\subsection{Extracting Dynamic Function}

Similar to static function extraction, downloading and unpacking are necessary to obtain dynamic functions.
Figure~\ref{fig:ext:dynamic} depicts how to collect dynamic functions in the sandbox when we install the software package.
The malicious packages may contain malicious codes, leading to computer system corruption, and we must separate the package installation process from the computer's regular procedures. 
We leverage the sandbox tool to isolate the package installation, which prevents malicious packages from exfiltrating sensitive data, accessing sensitive files (e.g., SSH keys), and persistent malware. 
Sandbox, in particular, builds a network firewall and an isolated filesystem layer by interposing on system calls (e.g., `open' and `connect') and re-writing system call arguments (e.g., file path).
During the execution of the software package, we utilize the STrace~\cite{strace} tool to capture its installation log file. 
For system administrators and security experts, a software package's logging file refers to the recording and documenting of when and how it was installed or executed. 
The logged system calls can reveal specific details, such as file I/O, process management, and network communication. 
The output can be extensive and supports multiple file formats for further analysis.

We use the regex matching to identify the dynamic functions in the logging file.
Table~\ref{tab:function} shows the overall information in three categories of static functions, including network-related, file-related, and process-related functions. 
If matched, we extract relevant information, including function names, arguments, and corresponding files.

\subsection{Findings and Lessons}

\begin{figure}[!t]
	\centering
	\includegraphics[width=2.4in]{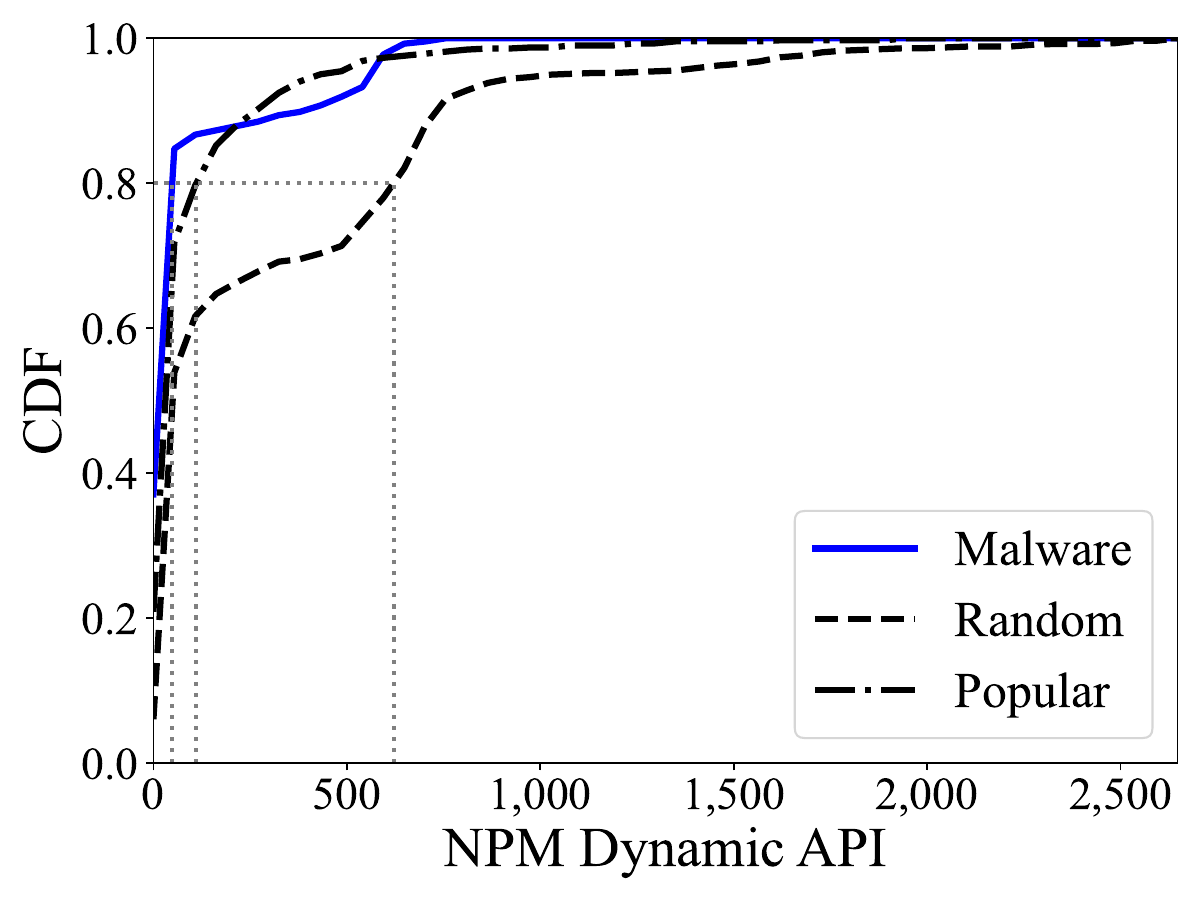}
	\caption{The CDF of dynamic functions in the NPM ecosystem.}
	\label{fig:npm:cdf}
\end{figure}

\begin{figure}[!t]
	\centering
	\includegraphics[width=2.4in]{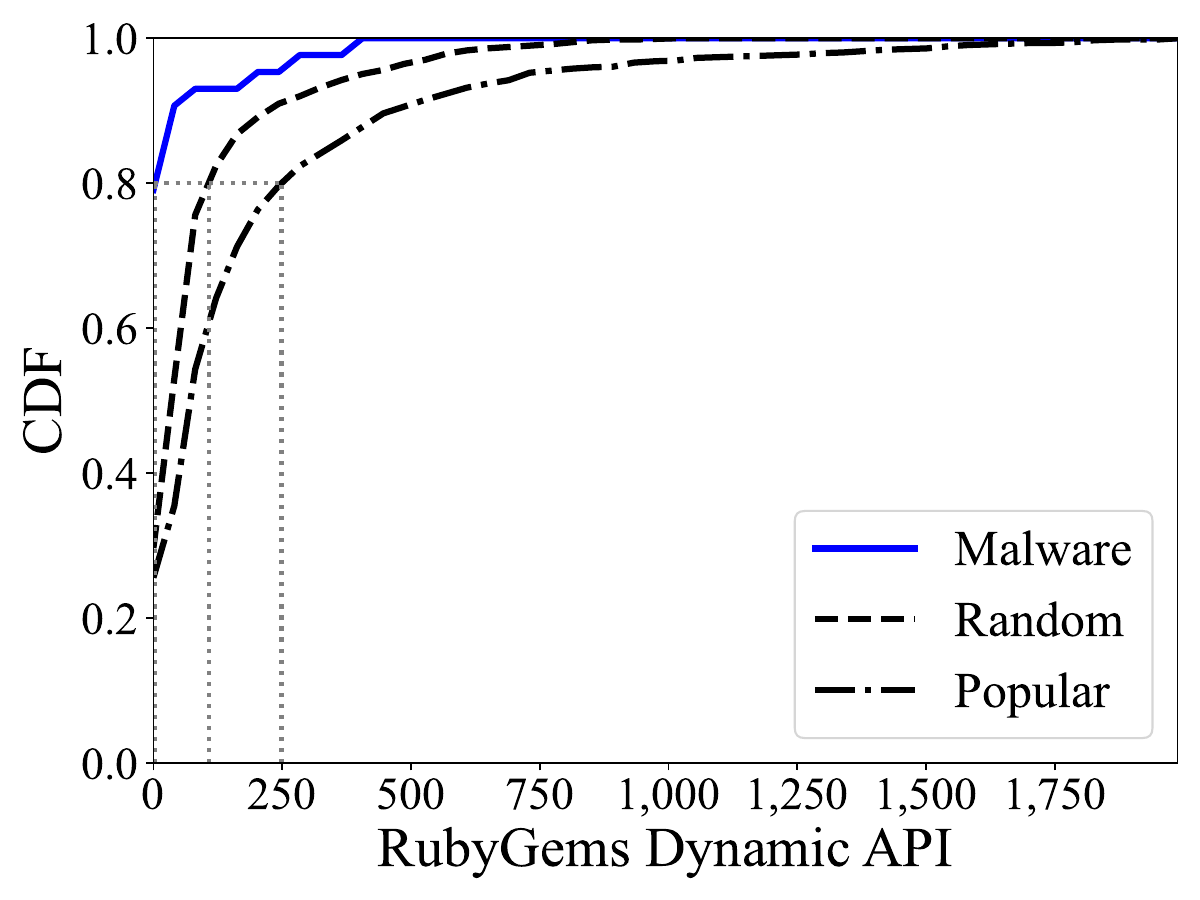}
	\caption{The CDF of the dynamic functions in the RubyGems ecosystem.}
	\label{fig:ruby:cdf}
\end{figure}

\textit{The number of dynamic functions}. 
We plot the CDF of dynamic functions in the NPM and RubyGems ecosystems. 
Figure~\ref{fig:npm:cdf} shows the distribution of dynamic functions of the NPM ecosystem, in which 80\% of the malicious packages have more than 2 functions in their logging trace. 
We observe that 80\% popular packages have more than 600 functions. 
Figure~\ref{fig:ruby:cdf} depicts the CDF of dynamic functions of the RubyGems ecosystem. 
Nearly 80\% malicious packages do not have any dynamic functions, while 80\% popular packages have a threshold of 250 functions.
The number of dynamic functions in legitimate functions is much larger than in malicious packages.
The plausible reason is malicious packages would like to install and run themselves in silent mode, avoiding being detected by the system.

We have several findings about the static and dynamic functions of software packages. 
First, the number of dynamic functions is much larger than that of static functions. 
The reason is that complex function call relationships exist between distinct files and libraries, resulting in multiple function calls in the dynamic trace. 
Second, the functions of popular packages surpass others (the malware and the random). 
It is due to the complexity of the functions and features of popular software packages that many functions are produced. 
Third, the malware has a long-tailed distribution of functions, with most packages having few function calls and a small number of packages having hundreds of function calls.

\begin{table}[!t]
    \centering
    \caption{The coefficient matrix of Legitimate and malicious packages in NPM}
    \begin{tabular}{c c c c}
    \toprule
     & file-network & file-process & network-process \\
    \midrule
    Legitimate & 0.17 & 0.37 & 0.19 \\
    Malicious & 0.10 & 0.61 & 0.05 \\
    \bottomrule
    \end{tabular}
    \label{Tab:Mal_Legi_coff}
\end{table}

\begin{figure*}[!t]
    \begin{tabular}{ c   c  c }
    \begin{minipage}[t]{0.31\linewidth}
    \centering
    \includegraphics[width=2.1in]{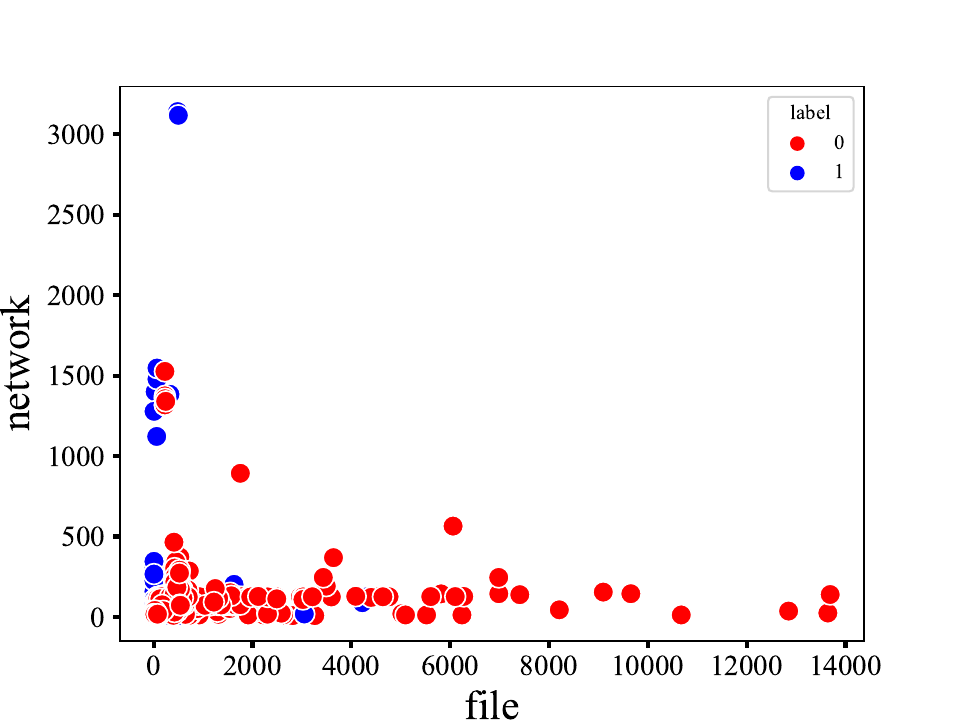}
	\caption{File-related and network-related functions.}
    \label{fig:f:n}
    \end{minipage}
    &
    \begin{minipage}[t]{0.31\linewidth}
        \includegraphics[width=2.1in]{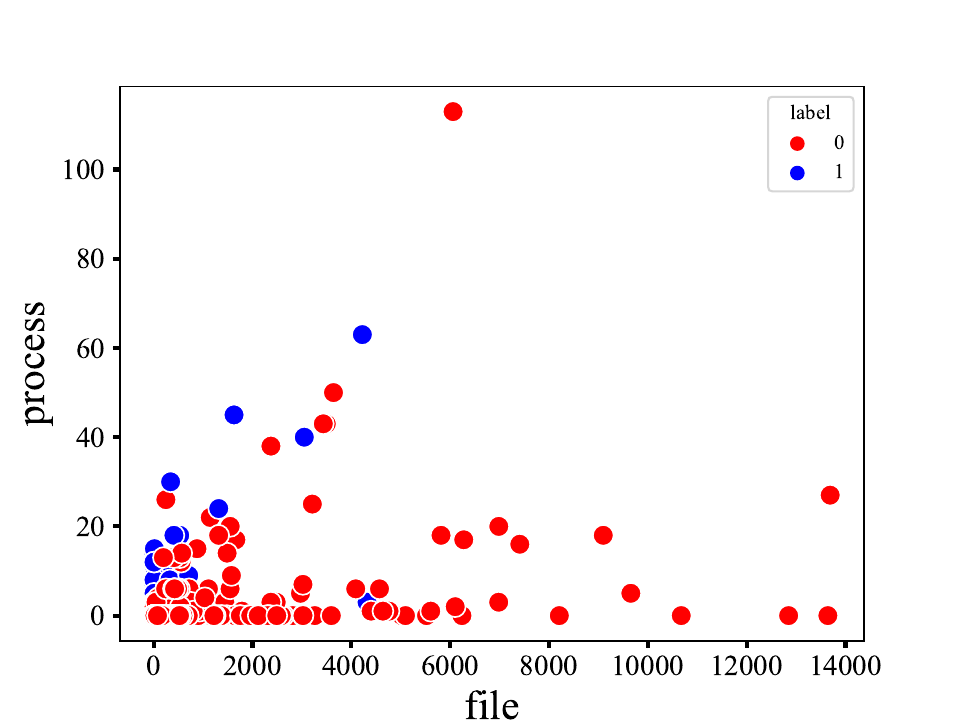}
        \caption{File-related and process-related functions.}
        \label{fig:f:p}
    \end{minipage}
	&
    \begin{minipage}[t]{0.31\linewidth}
        \includegraphics[width=2.1in]{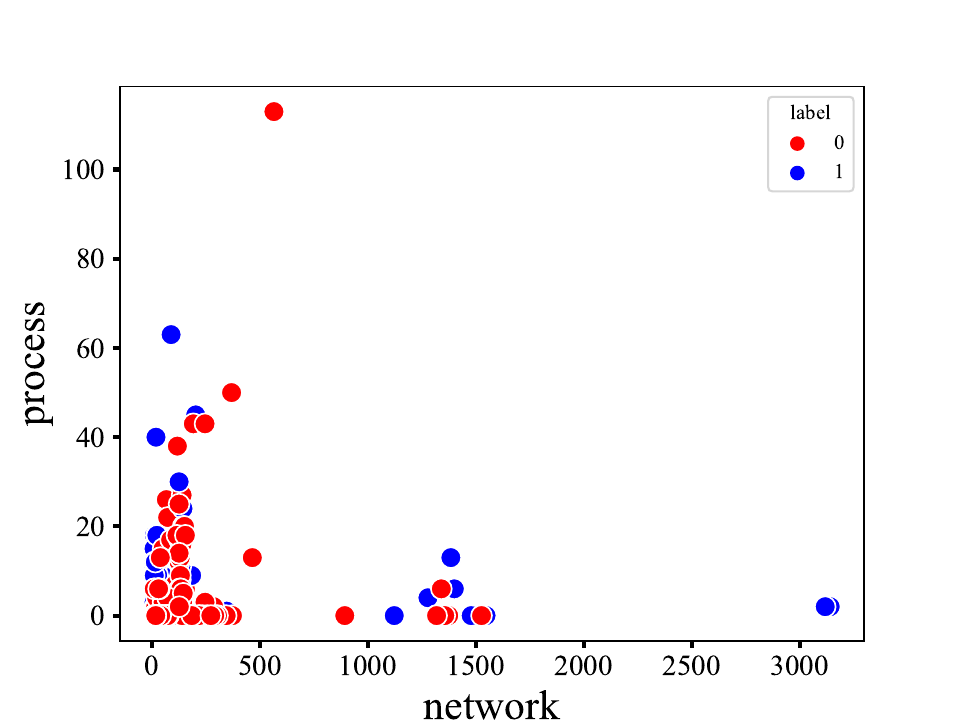}
        \caption{Network-related and process-related functions.}
        \label{fig:n:p}
    \end{minipage}
    \end{tabular}
\end{figure*}

\textit{Correlation Degree}.
We further provide an analysis of statistical relationships among the dynamic functions in three categories: network-related, file-related, and process-related functions. 
Specifically, we use Pearson correlation coefficient~\cite{zwillinger1999crc} to measure linear correlation between two sets of data.
Table~\ref{Tab:Mal_Legi_coff} lists Pearson correlation coefficients between different types of dynamic functions of legitimate and malicious packages. 
Pearson correlation coefficient ranges from -1 to 1, where -1 means a complete negative correlation, 1 indicates a complete positive correlation, and 0 means no linear correlation.
We further plot the overall statistics of dynamic APIs between legitimate and malicious packages in Figure \ref{fig:f:n}, \ref{fig:f:p}, \ref{fig:n:p}. 
The blue color point is the malicious package, and the red color point indicates the legitimate package.

First, combining Table~\ref{Tab:Mal_Legi_coff} and Figure~\ref{fig:f:n}, we can find that the correlation degree between file-related and network-related functions is small.
We find that most malicious packages have more network-related dynamic functions, while legitimate packages have more file-related.
This is consistent with the fact that many malware packages belong to C\&C servers, requiring lots of network requests.
Second, we observe a large correlation between file-related and process-related operations in the malicious package in Figure~\ref{fig:f:p} and Table~\ref{Tab:Mal_Legi_coff}. 
It indicates that a process-related operation may follow a file-related operation. 
Third, we observe the correlation between network-related and process-related operations in the malicious package is closed to zero in Table~\ref{Tab:Mal_Legi_coff}. 
Figure~\ref{fig:n:p} depicts that malicious packages stay around numerous network-related functions, while others have more process-related functions.

\textbf{Lessons} learned are following.
(1) The number of the dynamic function is a distinguishable indicator to distinguish between legitimate and malicious packages.
(2) Both legitimate and malicious packages call similar types of dynamic functions, where the content of dynamic functions may not be a good indicator for malware detection.
(3) Malicious packages have a high degree of correlation between file-related and process-related operations, indicating a pattern of malicious behavior.

%% file: R4.tex
\section{RQ4: Usage of FGI}
\label{sec:classifier}

In this section, we build a classification model to detect malicious packages based on fine-grained information (FGI).

\subsection{Embedding}

For the metadata information, we leverage word embedding to convert the sequence of words into numerical vectors.
The word embedding represents the metadata content and characteristics of textual information in a real-valued continuous vector space.
Here, we use Word2vec, which learns a low-dimensional vector to raw texts in the semantic form.
We have the embedding matrix $W_{corpus}$ $\in R^{d\times V}$, where $|V|$ is the vocabulary corpus, and $d$ is the real-valued word embedding vector.
Given a token ($word^i$), we convert it into the corresponding word embedding vector $v_{sem}^i$ through the matrix-vector product as follows,
\begin{equation}\label{equ:semantic}
  v_{meta}^i = W_{corpus}\cdot word^i
\end{equation}
In the vector, words that appear within similar contexts have similar vector representations.
Note that the embedding matrix $W_{corpus}$ is the parameter to be learned as the pre-trained model.
We use a two-layer neural network to learn the embedding matrix $W_{corpus}$.
Given the semantic content of the package, the token list $S_{meta} = \{word^1, word^2,..., word^n\}$ is converted into the vector list $S_{vec} = \{v^1, v^2, ..., v^n\}$, where the vector $v^i$ has the size $|d|$.

For the static/dynamic function information, we used lexical analysis techniques to tag segment the functions. 
This step consists of splitting the code of the function into basic syntactic elements (e.g., function names, parameters, and operators) for the subsequent vectorization.
After that, a sequence of functions is converted into a list $S_{f} = \{f^1, f^2,..., f^n\}$, where $f^i$ is one syntactic element of a function. 
Similarly, we use the embedding matrix $W_{f}$ to obtain the vector for the syntactic element $f^i$, as follows,
\begin{equation}\label{equ:api}
    v_{f}^i = W_{f}\cdot f^i
  \end{equation}
where $W_{f}$ is a pre-trained model for functions. 
We also use a two-layer neural network to learn the embedding matrix $W_{f}$.

\subsection{Classifier} 
A classifier refers to a mapping function between the input $x$ and the output $y$, denoted as $y=f(x)$, where the $x$ is the numerical vector and the $y$ is whether a package is malicious or legitimate. 
Here, we use two categories of learning algorithms to infer the classifier: (1) classic machine learning and (2) deep learning.

Specifically, we use the scikit-learn~\cite{scikit} library to implement 6 classic machine learning algorithms and the Pytorch~\cite{pytorch} library to implement two deep learning algorithms.
The classic machine learning algorithms include Decision Tree (DT), Linear Regression (LR), Support Vector Machine (SVM), Random Forest (RF), k-nearest Neighbors (KNN), and Neural Network (NN).
The deep learning algorithms include Long Short-Term Memory (LSTM) and Convolutional Neural Network (CNN).
Table~\ref{tab:parameter} lists parameters of the eight learning algorithms, where we utilized grid cross-validation to find their optimal parameters.

\begin{table}[!t] \small
	\caption{Experimental parameters.}
	\centering
	\begin{tabular}{c   c  }
		\toprule
		                       &Parameters \\
		\toprule    
        DT    &  \makecell[c]{max-depth: 10, min-samples-split: 5 }      \\
		LR & C: 0.1    \\
        SVM & C: 0.1, kernel: linear     \\
		RF & \makecell[c]{min-samples-split: 2, max-depth: 20, n-estimators: 200    }    \\
        KNN & n-neighbors: 5      \\
		NN   & hidden-layer-sizes: (100,)           \\ \hline
        LSTM & \makecell[c] {batch-size: 32, epochs: 20, filters: 32, kernel-size: 3}         \\
        CNN & \makecell[c] {batch-size: 32, epochs: 20,  filters: 64, kernel-size: 3}\\
		\toprule
	\end{tabular}
    \label{tab:parameter}
\end{table}

\begin{table}[!t] \small
	\caption{Performance: the metadata information.}
	\centering
	\begin{tabular}{c  c c c c }
		\toprule
		             &    Accuracy 
                         &  Precision 
                         &  Recall   
                         &  F1-score\\
		\toprule    
        DT   &87.9\%           &    85.9\%      &    \textbf{88.7\%}   &       87.3\% \\
		LR &   89.8\%           &    92.2\%      &      85.5\%      &       88.7\% \\
        SVM     &   90.9\%           &    93.9\%      &      86.3 \%       &       90.0\% \\
		RF &   \textbf{92.8\%}           &    \textbf{96.4\%}      &   87.9\%    &       \textbf{92.0\%}  \\
        KNN     &   88.3\%           &    88.4\%      &      86.3\%       &       87.3\% \\ 
		NN          &   91.3\%           &    93.2\%      &      87.9\%      &       90.4\% \\\hline
        LSTM  &           81.9\% &      84.5\% & 75.0\% & 79.5\%\\
        CNN & 92.5\% & 94.8\% & \textbf{88.7\%} & 91.7\%\\
		\toprule
	\end{tabular}
    \label{tab:model:metadata}
\end{table}

\subsection{Performance}

We have conducted experiments to validate the performance of the classifier.
We pick up the NPM ecosystem from Table~\ref{tab:data} as the dataset for learning the classifier, including 686 malicious and 2,000 legitimate packages.
We selected 80\% of the dataset as the training data and 20\% as the test set.
Here, we use four metrics to represent the model's performance: accuracy, precision, recall, and F1-score. 
The precision is equal to $(TP)/(TP+FP)$; the recall is equal to $(TP)/(TP+FN)$; the accuracy is equal to $(TP+TN)/(P+N)$; and the F1-score is the harmonic mean of precision and recall.

\textbf{Metadata}. 
Table~\ref{tab:model:metadata} lists the performance of 8 algorithms based on the metadata information of the software package. 
It is obvious that the metadata information plays a useful indicator in distinguishing legitimate and malicious packages, consistent with our analysis in Section~\ref{sec:metadata}.
Classic machine learning algorithms (DT, LR, SVM, RF, KNN, and NN) perform similarly to deep learning algorithms (LSTM and CNN). 
The RF achieves the best performance among all algorithms. 
The average performance achieves 90\% accuracy, 88\% precision, 84\% recall, and 83\% F1-score.
The plausible reason is that the metadata belongs to a distinguishable pattern, and the embedding approach can find meaningful features in malware detection.

\begin{table}[!t] \small
	\caption{ Performance: Static Functions.}
	\centering
	\begin{tabular}{c    c c c c }
		\toprule
		                 &    Accuracy 
                         &  Precision 
                         &  Recall   
                         &  F1-score\\
		\toprule    
        DT    &81.3\%           &    73.3\%      &    \textbf{91.7\%}   &       81.5\% \\
		LR    &   \textbf{91.3\%}           &    89.2\%      &      \textbf{91.7\%}      &       90.4\% \\
        SVM    &   87.0\%           &    93.6\%      &      81.5 \%       &       87.1\% \\
		RF  &   88.8\%           &    84.6\%      &   \textbf{91.7\%}    &       88.0\%  \\
        KNN      &   87.5\%           &    86.1\%      &      86.1\%       &       86.1\% \\ 
		NN            &   90.0\%           &    86.8\%      &      \textbf{91.7\%}      &       89.2\% \\\hline
        LSTM  &           83.8\% &        92.6\% & 69.4\% & 79.4\%\\
        CNN  & \textbf{91.3\%} & \textbf{96.8\%} & 83.3\% & \textbf{89.6\%}\\
		\toprule
	\end{tabular}
    \label{tab:model:static}
\end{table}

\textbf{Static Function}. 
Table~\ref{tab:model:static} lists the performance of 8 algorithms based on the static function of the software package. 
Hence, we remove the packages with the empty static function.
The average performance achieves 86\% accuracy, 84\% precision, 79\% recall, and 82\% F1-score.
The static function also plays a positively correlated relationship between legitimate and malicious packages.
The static function is slightly inferior to the metadata.
The CNN achieves the best performance, and the other 5 algorithms (DT, LR, SVM, KNN, and NN) perform similarly. 
By contrast, the LSTM has the worst performance because it tends to capture patterns in data sequences, e.g., in the context of natural language processing. 
Yet, the static function extraction drops sequence relationships.

\begin{table}[!t] \small
	\caption{Performance: Dynamic Functions.}
	\centering
	\begin{tabular}{c c c c c }
		\toprule
		                     &    Accuracy 
                         &  Precision 
                         &  Recall   
                         &  F1-score\\
		\toprule    
        DT   &81.9\%           &    74.1\%      &    94.4\%   &       83.0\% \\
		LR    &   84.9\%           &    77.6\%      &      \textbf{95.2\%}      &       85.5\% \\
        SVM     &  \textbf{85.3\%}           &    78.5\%      &      94.4 \%       &      \textbf{85.7\%} \\
		RF   &   83.8\%           &    77.2\%      &   92.7\%    &       84.2\%  \\
        KNN     &   84.9\%           &    \textbf{98.8\%}      &      68.5\%       &       81.0\% \\ 
		NN            &   84.2\%           &    77.3\%      &      93.5\%      &       84.7\% \\ \hline
        LSTM &           78.5\% &    72.8\% & 86.3\% & 79.0\%\\
        CNN & 83.8\% & 76.8\% & 93.5\% & 84.4\%\\
		\toprule
	\end{tabular}
    \label{tab:model:dynamic}
\end{table}

\begin{table}[!t] \small
	\caption{Performance: Metadata + Static + Dynamic functions.}
	\centering
	\begin{tabular}{c   c c c c }
		\toprule
		                        &    Accuracy 
                         &  Precision 
                         &  Recall   
                         &  F1-score\\
		\toprule    
        DT   &91.3\%           &    89.1\%      &    92.7\%   &       90.9\% \\
		LR    &   97.0\%           &    97.5\%      &      96.0\%      &       \textbf{96.7\%} \\
        SVM     &   \textbf{96.2\%}           &    95.2\%      &      \textbf{96.8 \%}       &       96.0\% \\
		RF   &   94.7\%           &   \textbf{ 98.2\% }     &   90.3\%    &       94.1\%  \\
        KNN   &   91.7\%           &    88.1\%      &      95.2\%       &       91.5\% \\ 
		NN      &   95.8\%           &    97.5\%      &      93.5\%      &       95.5\% \\ \hline
        LSTM &           84.5\% &        95.6\% & 70.2\% & 80.9\%\\
        CNN & \textbf{96.2\%} & 97.5\% & 94.4\% & 95.9\%\\
		\toprule
	\end{tabular}
    \label{tab:model:all}
\end{table}

\textbf{Dynamic Function}. 
Table~\ref{tab:model:dynamic} lists the performance of 8 algorithms based on the dynamic function of the software package.  
The average performance achieves 82\% accuracy, 78\% precision, 86\% recall, and 82\% F1-score.
Note that dynamic functions are extracted in the logging file when the system runs a software package.
Most dynamic functions are affected by various factors, such as network conditions and the runtime state of systems.
The dynamic function is slightly inferior to the metadata.
Except for LSTM, most algorithms have similar performance.

\textbf{Metadata + Static + Dynamic Function}.  
We directly concatenate those vectors ($v_{sem}$ and $v_{f}$) into one vector as [$v_{sem}$, $v_{f}$].
Table~\ref{tab:model:all} lists the performance of eight algorithms based on all FGI elements.
The concatenation of 3-dimensional information archives the best performance, 95\% accuracy, 93\% precision, 90\% recall, and 91\% F1-score.  
We compared the different performance between Tables(\ref{tab:model:metadata} \ref{tab:model:static} \ref{tab:model:dynamic}) and Table~\ref{tab:model:all}, and found that the combined FGI has a slight performance improvement. 
There are two possible reasons. 
First, the metadata belongs to the natural language processing domain, and the function belongs to the source code domain, leading to the multi-modal information. 
Different modalities have fundamentally different statistical properties and patterns.
A combination of various modalities of data is non-trivial.	
Second, those erroneous predictions are caused by function extraction failure and similar packages. 
The unbalanced data distribution also affects the classification performance.

\textbf{Lessons} learned are summarized below.
(1) The FGI is a distinguishable indicator for detecting malicious packages. 
(2) The metadata performs better than static/dynamic functions in malware detection. 
However, the drawback of metadata is that attackers can easily change its content, making the malware detection approach invalid.
(3) One-dimensional information has sufficient distinguishable capability to detect a malicious package. 
Simply combining different dimensional information would not significantly improve the overall performance.

%% file: discuss.tex
\section{Analysis Validity}

To guarantee the reproducibility of our analysis results, we follow the prudent experimental requirements.

\textbf{Result Transparency}.
We provide the details of the transparency of the package dataset, including 50,000+ legitimate and 1,000 malicious packages. 
Due to the page limits, we present only a summary of our dataset in the paper. The details of the dataset are available in the GitHub repository.
We build a website to publish all package names (sources) with their signatures (e.g., MD5 hashes) in our paper.

\textbf{Result Correctness}. 
For the dataset correctness, we use a heuristic rule to filter out false positives. 
If a legitimate package is falsely reported as a malicious package, it may be falsely labeled as such.
If a package is not removed by the root register, it is not malicious, and we remove it. 
For the correctness of the package FGI, we have manually inspected the FGI content to guarantee the correctness of experimental results. 
Further, we release our dataset via the GitHub repository, where we can receive feedback from the community to remove the false positives.

\textbf{Stability Issue}.
One concern is that the analysis results may change when various legitimate packages are added, or new malicious packages are added. 
First, our legitimate packages are chosen from popular packages (the most downloaded or highest Pagerank) and random packages.
Second, we survey the release time of legitimate/malicious packages from 2008 to 2022. Our dataset covers an extended period, and the analysis results are stable with time.

%% file: relate.tex
\section{Related Work}

\textbf{OSS Ecosystems}.
Millions of packages have been released in OSS ecosystems, and various research works~\cite{kabbedijk2011steering, german2013evolution, serebrenik2015challenges, ma2018constructing, dey2018software, ma2019world, pashchenko2020preliminary} on package measurement and analysis have been conducted.
\citet{decan2017empirical} found that packages in OSS ecosystems have become more complicated as time passed, and the dependency numbers have increased. 
In addition, they pointed out that conflicted versions and package compatibility are the major impeding causes of deprecated, redundancy, and dependency relationships.
\citet{constantinou2017empirical} compared developer retention between the RubyGems and NPM ecosystems, where many software packages lack maintenance.
\citet{kikas2017structure} studied the evolution of dependencies and the vulnerability of the dependency network in the NPM ecosystem.
\citet{zimmermann2019small} downloaded all versions of all published packages with several snapshots and studied several security risks in the NPM ecosystem, including direct and transitive dependency concerns. 
\citet{pashchenko2020qualitative} claimed that many developers are unaware of dependency issues or do not attempt to modify the software. When there is a vulnerability in a dependent package, developers do not update the fixed package version promptly, resulting in the transmission of the vulnerability. 
\citet{zahan2021weak} analyzed the metadata of 1.63 million packages and provided six indicators of possible security risks in NPM.

\textbf{Software Package Security}.
As the ecosystem grows, the package security risk increases, which could be exploited as a launching pad by attackers.
\citet{decan2018evolution,decan2018impact} studied nearly 400 security reports of the NPM software packages and found that the number of vulnerable packages constantly increases.
\citet{vaidya2019security} pointed out that private information is leaked in the code of software packages, including critical files and API keys embedded in the code.
\citet{xiao2021abusing} proposed an attack that abuses hidden attributes, which attackers can exploit to obtain confidential data, bypass security checks, and launch denial-of-service attacks. 
\citet{alfadel2021empirical} studied a collection of 550 vulnerabilities affecting 252 PyPi packages, and their analysis indicated that vulnerabilities grew over time, and the most common was XSS vulnerabilities.
\citet{ponta2018beyond} proposed a code-centric scheme for detecting, analyzing, and mitigating vulnerabilities in software packages. 
\citet{woo2021v0finder} proposed the V0Finder to discover the original software vulnerabilities by integrating diverse data sources and utilizing machine learning techniques.
Sejfia et al.~\cite{SejiaAdria2022Machinelearn} presented a machine-learning-based approach for automatically detecting potentially malicious packages.
There are prior works~\cite{reif2016call, patra2018conflictjs, lauinger2018thou, robbie2012API, wang2020watchman, li2021arbitrar} to find the API calls in a software package.
Further, the research community has paid attention to OSS malware, including malware detection~\cite{cappos2008look, SejiaAdria2022Machinelearn, zhang2020cyber, qian2022malicious, ferreira2021containing, vu2023bad}, malware analysis~\cite{wy2022InstalltimeAtt, lidisa2022javabytecode, sejfia2022practical}, and malware data collection~\cite{backstabbers-online, Ohmssc2020, guo2023empirical}.
By contrast, our analysis compares legitimate and malicious packages in three granularity levels.

%% file: conclusion.tex
\section{Conclusion}
This paper presents a large-scale study of the fine-grained information extraction and analysis of software packages covering 3 OSS ecosystems. 
Our investigation covers 50,000 legitimate and 1,000 malicious packages, each divided into 3 levels of FGI: metadata, static, and dynamic functions.
Our comparison reveals several findings.
First, the legitimate's FGI differs greatly from the malware's FGI.
Malicious packages have less metadata and employ fewer static/dynamic functions than legitimate packages.
Second, legitimate and malware have different tendencies in terms of call functions, e.g., malicious packages are prone to use HTTP/URL functions rather than FTP or SMTP. 
Third, the detection approach based on FGI achieves promising performance in distinguishing legitimate and malicious packages.
Fourth, one-dimensional FGI has sufficient distinguishable capability to detect malicious packages, and simply combining different dimensional FGI cannot improve overall performance.
